\documentclass{aa}

\usepackage[colorlinks=true,citecolor=blue,urlcolor=blue]{hyperref}

\usepackage{graphicx}
\usepackage{txfonts}
\usepackage{lipsum}
\usepackage{subcaption}        
\usepackage{lscape}             
\usepackage{placeins}          
\usepackage{threeparttable}
\newcommand{\teff}{$T_{\mathrm{eff}}$}

\begin{document}

   \title{Tracing the early Milky Way thin disc with the Gaia-ESO Survey}

   \subtitle{}

   \author{C. Viscasillas Vázquez \inst{1,2} 
   \and
   S. Randich\inst{2} 
   \and R. E. Giribaldi\inst{2}\
   \and L. Berni\inst{2,3} 
   \and L. Magrini\inst{2} 
   \and M. Tsantaki\inst{2} 
        }

   \institute{Institute of Theoretical Physics and Astronomy, Faculty of Physics, Vilnius University, Sauletekio av. 3, 10257 Vilnius, Lithuania.
   \and INAF – Osservatorio Astrofisico di Arcetri, Largo E. Fermi 5, 50125 Firenze, Italy.
   \and Dipartimento di Fisica e Astronomia, Università degli Studi di Firenze, Via Sansone 1, 50019, Sesto Fiorentino, Italy.}

    \date{Received 22 June 2026 / Accepted 6 August 2026}

  \abstract
  {The origin of metal-poor stars on thin-disc-like orbits remains an open question in Galactic archaeology and provides important constraints on the earliest phases of Milky Way disc formation. While low-metallicity stars are usually associated with the stellar halo or the thick disc, recent large spectroscopic surveys have revealed stars with low metallicities on cold, prograde, nearly circular orbits. These objects challenge the classical view of the thin disc as a young and metal-rich component, and may trace the low-metallicity tail of an old thin disc, the high-angular-momentum tail of the metal-weak thick disc, or prograde halo-related populations.}
 {We aim to identify and characterise metal-poor stars with thin-disc-like kinematics observed in the Gaia-ESO Survey. We investigate their dynamical, chemical, and evolutionary properties in order to assess whether they can be associated with the old, metal-poor tail of the Galactic thin disc.} 
 {We use orbital parameters derived from {\em Gaia} DR3 data, combined with spectroscopic information from the Gaia-ESO Survey. Stellar ages are obtained through isochrone fitting, while detailed chemical abundances for several elements and kinematics are used to place the stars in the context of Galactic populations.}
 {Out of 1784 turn-off stars, we identify one metal-poor candidate star with thin-disc-like kinematics. It has $\mathrm{[Fe/H]}=-1.38$ and follows a dynamically cold, prograde orbit, with low eccentricity, high azimuthal velocity ($V_{\phi} \simeq 220,\mathrm{km~s^{-1}}$), large angular momentum, low vertical action, and a guiding radius close to the solar neighbourhood. Chemically, however, it is clearly distinct from the thin disc: it is $\alpha$-enhanced ($[\mathrm{Mg/Fe}] = +0.58$; [$\alpha$/Fe]= +0.48), shows a low $[\mathrm{Y/Mg}] = -0.58$ ratio, and its multi-element abundance pattern is offset from that of the thin-disc reference. Its position in the $[\mathrm{Al}/\mathrm{Fe}]$--$[\mathrm{Mg}/\mathrm{Mn}]$ plane further confirms that its chemistry is distinct from the thin-disc sequence. The star also has an old age estimate, with a most probable value of $\sim 10.2~\mathrm{Gyr}$ and a broad 1$\sigma$-like interval of $\sim4.4$--$13.4~\mathrm{Gyr}$.}
 {The combination of low metallicity, old age, $\alpha$ enhancement, low $[\mathrm{Y/Mg}]$, a position on the metal-poor sequence in the $[\mathrm{Al}/\mathrm{Fe}]$--$[\mathrm{Mg}/\mathrm{Mn}]$ plane, and dynamically cold thin-disc-like kinematics suggests that this star is not a typical member of the canonical thin disc. Instead, it is a strong candidate for a metal-poor relic of the early Galactic disc, possibly associated with the proto-thin disc or with the transition between the proto-Galactic halo and the emerging disc. Its properties show that stars with halo-like metallicities and chemically distinct abundance patterns can occupy orbits characteristic of the thin disc, highlighting the existence of metal-poor stars on dynamically cold orbits while emphasizing that their origin may involve the old thin disc, the metal-weak thick disc, or prograde halo-related populations. This object provides a useful benchmark for studying the chemical enrichment and dynamical state of early metal-poor populations on disc-like orbits.}


   \maketitle
   \nolinenumbers

\section{Introduction}

Metal-poor stars preserve crucial information on the earliest phases of the Milky Way (MW) assembly. 
For a long time, such stars were mainly associated with the stellar halo, while the Galactic disc, and in particular the thin disc, was thought to be a predominantly more metal-rich component. This classical view has recently been challenged by the discovery of stars with low (metal poor, MP, [Fe/H]<-1.0) and very low metallicities (very metal poor, VMP, [Fe/H]<-2.0) that can also be found on disc-like orbits, reopening the question of when the MW disc first formed and what its earliest stellar populations looked like \citep[e.g.][]{Sestito2019,Sestito2020,Cordoni2021,Carter2021,Carollo2023,FernandezAlvar2021,FernandezAlvar2024,ArdernArentsen2024}. In particular, recent {\em Gaia}-based analyses have strengthened the evidence of low-metallicity stars with disc-like kinematics and a preference for prograde orbits. Using a large local sample with photometric metallicities and accurate {\em Gaia} DR3 astrometry, \citet{Bellazzini2024} showed that stars with [Fe/H] $\leq -1.5$ and disc-like orbital properties are significantly more common on prograde than on retrograde orbits, and argued that part of this population may trace an early in situ disc component. In parallel, chrono-chemo-dynamical analyses of {\em Gaia} Radial Velocity Spectrometer (RVS) samples have suggested that the MW may host an extremely old thin disc, with ages extending beyond 13 Gyr, thus implying that a dynamically cold disc could have been in place within the first Gyr after the Big Bang \citep{Nepal2024}. These results are particularly intriguing in the broader context of high-redshift observations, which have revealed surprisingly early disc-like structures in distant galaxies \citep[e.g.][]{Neeleman2020,Rizzo2020,Ferreira2022,Kartaltepe2023,Robertson2023}. 

However, the interpretation of metal-poor stars on present-day disc-like orbits is not unique. While some may be genuine relics of an early Galactic disc or members of the metal-poor tail of the old thin disc, others could belong to the low-metallicity tail of the thick disc or the metal-weak thick disc (MWTD) 
\citep[e.g.][]{Norris1985,Chiba2000,Beers2002,Ruchti2010,Beers2014}. The MWTD may therefore be regarded as a low-metallicity, thick-disc-like component, potentially overlapping with the metal-poor tail of the canonical thick disc \citep[e.g.][]{Kordopatis2013}. Alternatively, some of these stars may represent objects accreted on low-inclination orbits or early disc relics later affected by dynamical heating and merger events 
\citep[e.g.][]{Santistevan2021,Sestito2021,SotilloRamos2023,McCluskey2024,Sestito2026}. This ambiguity is particularly important because neither the thick disc nor the halo is dynamically or chemically homogeneous; for instance, \citet{Zhang2024} showed that a rotation-supported disc population begins to emerge around $[\mathrm{M/H}]\sim -1.3$, while at lower metallicities prograde halo components may explain part of the apparent excess of stars on disc-like orbits. Moreover, for very old stars, present-day orbital parameters may not reflect birth conditions, since radial migration, bar/spiral interactions, changes in the Galactic angular-momentum axis, and merger-driven heating can alter angular momentum and orbital structure over time 
\citep[see e.g.][]{Sellwood2002,minchev2010,Aumer2017,Dillamore2022,Yuan2024}.

Despite this progress, the nature of metal-poor stars on disc-like orbits remains debated. High-quality spectroscopy is therefore essential to assess the nature of individual candidates through their detailed atmospheric parameters and chemical-abundance patterns. In this work, we report the identification of a metal-poor star with thin disc-like orbital properties in the Gaia-ESO Survey \citep[GES, ][]{Gilmore2022,Randich2022}. The finding of such an object in the relatively small GES sample provides an important high-resolution spectroscopic counterpart, to the populations recently identified by \citep{FernandezAlvar2024,Nepal2024}; this star offers a uniquely detailed chemical inventory, enabling a more comprehensive assessment of its origin. We therefore examine its kinematic and chemical properties and assess whether it may represent a plausible relic of the early MW disc. The paper is organised as follows: Section~\ref{sec:data} describes the data and sample selection. Section~\ref{sec:kinematics} presents the kinematic and stellar properties of the candidate, including its orbital parameters and age estimate. Section~\ref{sec:chemistry} analyses its detailed chemical properties. Section~\ref{sec:discussion} discusses the possible origin of the star in the context of metal-poor disc populations. Finally, Section~\ref{sec:summary} summarises our main conclusions.

\section{Data and sample selection}
\label{sec:data}

Our analysis is based on high-resolution spectroscopic data from the Gaia-ESO Survey, specifically from the UVES sample (\texttt{SETUP} = \texttt{U580}). Within the GES classification scheme, we restricted our analysis to targets observed as part of the Milky Way field stars (\texttt{GES\_FLD} = \texttt{GE\_MW}), excluding calibration fields, star cluster, and benchmark stars. This selection results in a sample of 3298 stars. We further focused on stars located around the main-sequence turn-off (MSTO), as their age estimate is more reliable \citep[e.g.][]{Howes2019}.
To perform this selection, we adopted cuts in surface gravity and effective temperature, selecting stars with $3.8 \leq \log g \leq 4.3$ and $5600 \leq T_{\rm eff} \leq 6900$ K, following previous works \citep[e.g.][]{Chen2022,Viscasillas2023}. This selection yields a final sample of 1784 stars.

The orbital parameters were taken from the catalogue of \citet{Kordopatis2023}, where the orbits were computed using {\em Gaia} astrometry, {\em Gaia} line-of-sight radial velocities, and the geometric and photogeometric distances of \citet{BailerJones2021}. The adopted solar position is $(R,Z)_\odot=(8.249,0.0208)$ kpc, and the solar Galactocentric velocity is $(V_R,V_\phi,V_Z)_\odot=(-9.5,250.7,8.56)$ km s$^{-1}$. Orbital parameters, including eccentricity, angular momentum, and the maximum height above the Galactic plane, were then derived with the Stäckel fudge method using \textsc{galpy} and the axisymmetric Galactic potential of \citet{McMillan2017}. For our sample, the {\em Gaia} and Gaia-ESO radial velocities ($\mathrm{RV}$) show good overall agreement, with a median difference of $\Delta \mathrm{RV} = \mathrm{RV}_{\rm Gaia} - \mathrm{RV}_{\rm GES} = +0.29$ km s$^{-1}$ and a 16th--84th percentile range of $[-3.32,+4.44]$ km s$^{-1}$. This is consistent with the comparative analysis of Gaia-ESO DR5.1 and Gaia DR3 GSP-Spec by \citet{VanDerSwaelmen2024}, who found excellent agreement between the Gaia and Gaia-ESO radial velocities within the uncertainties of each dataset. This indicates that the use of the {\em Gaia}-based orbital parameters does not introduce a significant systematic bias in our kinematic interpretation.
The derived orbital parameters are consistent with those obtained using independent methodologies, such as the calculations presented by \citet{Berni2025}, who computed stellar orbits assuming the \citet{McMillan2017} Galactic potential using the \texttt{AGAMA} code \citep{Vasiliev2019} and following the prescriptions described in \citet{Massari2019}. In addition, for the purpose of constructing the Lindblad diagram, we computed the orbital energy for each star following the same framework, using the results of \citet{Berni2025}. We also derived the Galactic velocity components in the local standard of rest ($U_{\rm LSR}, V_{\rm LSR}, W_{\rm LSR}$), which are used to construct the Toomre diagram. The guiding radii $R_{\rm guid}$ were computed with  \textsc{galpy} \citep{Bovy2015}, adopting the same Galactic potential used for the orbital integration.

Stellar ages were derived via fitting Yale-Yonsey isochrones \citep{yi2001ApJS..136..417Y,kim2002,yi2003} to the inputs effective temperature (\teff), [Fe/H], and absolute visual magnitude ($M_V$) adopting the frequentist approach implemented in the Python code \texttt{q2} \citep{Ramirez2014}.
The algorithm incorporates solar-scaled isochrones, spaced in steps of 0.02~dex in [Fe/H].
The effects of $\alpha$-enhancement are accounted following the strategy in \cite{spina2018MNRAS.474.2580S}, which includes in the metallicity the effect of [$\alpha$/Fe] according to the relation of \cite{salaris1993ApJ...414..580S}:
\begin{equation}
\label{eq:alpha_enh}
    [\rm{M/H}] = [\rm{Fe/H}] + \rm{log}(0.638 \times 10^{[\alpha/Fe]} + 0.362)
\end{equation}

\noindent
where $\alpha$ is the average of Mg, Si, Ca, and Ti;  see details of the effect of the $\alpha$ enhancement in Appendix~\ref{app:alpha}. 
 The availability of these abundances limited our selected sample of 1784 stars to 1263; the method converged for 1109 stars from that input data.
We adopted the zero-point offset of $-0.04$~dex in [Fe/H], attributed to atomic diffusion \citep{melendez2012A&A...543A..29M,dotter2017ApJ...840...99D,spina2018MNRAS.474.2580S}, with which the age of the Sun (4.6~Gyr) is recovered from its canonical parameters.
The method internally computes $M_V$ from the  extinction corrected magnitude $V_0$ and the distance. For the latter we adopted the geometric distances of \cite{bailer-jones2021AJ....161..147B}, which are based on Gaia parallaxes and are corrected for its zero-point offset \citep{Lindegren2021A&A...649A...4L}. 
$V_0$ was computed adopting the reddening estimates in Appendix~\ref{app:reddening}.

To compute the ages, \teff\ and [Fe/H] were adopted from the Gaia-ESO catalogue, and their errors are assumed to be of 100~K and 0.1~dex, respectively. For $M_V$, we assumed an error of 0.05~mag, obtained by adding in quadrature the  error of the extinction  in the $V$ band  $3.1 \times \sigma(E(B-V)) = 3.1 \times 0.013 = 0.04$~mag --estimated in Appendix~\ref{app:reddening}-- and a conservative error of the Gaia-to-Johnson transformation of $\sigma_V = 0.03$~mag.

\section{Kinematic and stellar properties}
\label{sec:kinematics}

Dynamical and age information are fundamental for constraining the origin of stellar populations. In the following, the kinematic and orbital properties are discussed first, followed by an analysis of the age and evolutionary status.

\subsection{Kinematic identification and orbital properties}
\label{subsec:kin}

The aim of this work is to select metal-poor stars ([Fe/H] < -1) with orbital properties consistent with thin disc populations. Therefore, our first selection is based on orbital parameters, including angular momentum, azimuthal velocity, eccentricity, and orbital energy. Among the stars exhibiting thin-disc-like kinematics-namely low orbital eccentricities, high rotational support, and small vertical excursions from the Galactic plane, and with [Fe/H] < -1, we select one candidate (CNAME = 15183399$-$0721310). Its stellar and orbital parameters are summarised in Table~\ref{tab:target}. The Gaia-ESO radial velocity of the candidate is in excellent agreement with the Gaia DR3 value used for the orbital determination, with $\mathrm{RV}_{\rm GES}=-28.81$ km s$^{-1}$ and $\mathrm{RV}_{\rm Gaia}=-27.81$ km s$^{-1}$.

\begin{figure}
    \centering
    \includegraphics[width=\columnwidth]{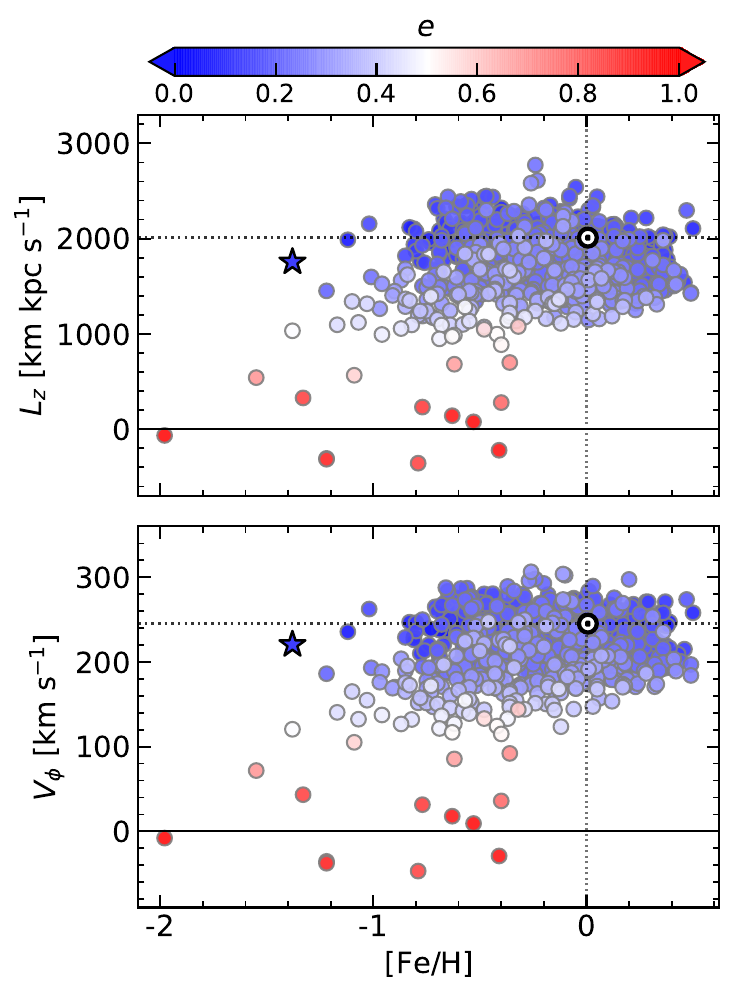}
    \caption{Distribution of the MSTO sample in the $L_z$--[Fe/H] (top) and $V_{\phi}$--[Fe/H] (bottom) planes, colour-coded by orbital eccentricity. The star symbol corresponds to the identified candidate, while the solar symbol marks the position of the Sun for reference.}
    \label{fig:feh_kinematics}
\end{figure}

\begin{figure}
    \centering
    \includegraphics[width=\columnwidth]{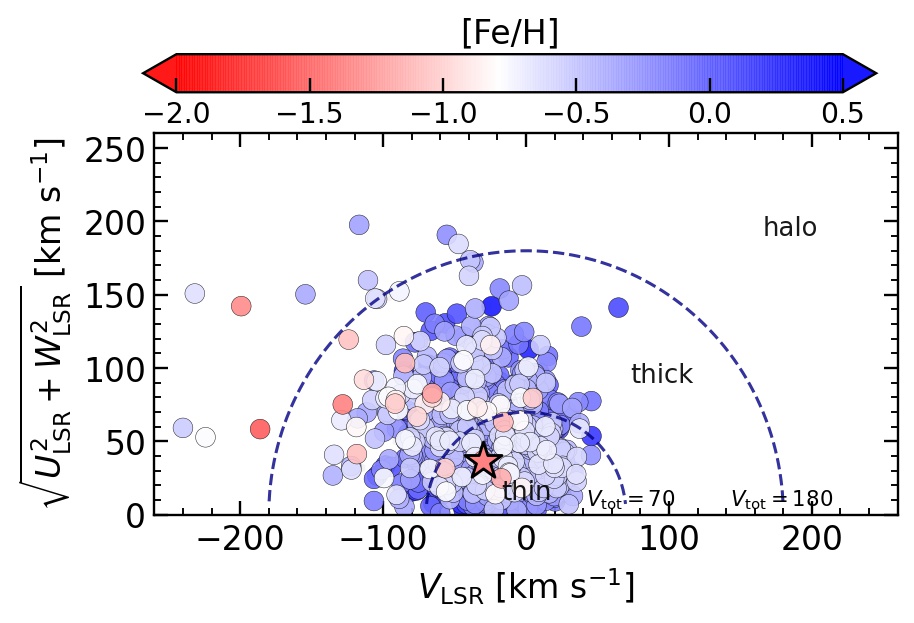}
    \caption{Toomre diagram for the MSTO sample colour-coded by metallicity. The candidate star is highlighted with a star symbol. The dashed curves indicate the kinematic loci commonly used to distinguish thin-disc, thick-disc, and halo populations based on total velocity with respect to the LSR.}
    \label{fig:toomre}
\end{figure}

The candidate has a metallicity of $\mathrm{[Fe/H]} = -1.38 \pm 0.06$, below the classical lower limit for the thin disc [Fe/H]$=-0.7$ \citep{Bensby2014} or even the recently proposed value by \cite{Hu2025} ([Fe/H]$=-1.2$). However, its kinematic properties clearly distinguish it from the bulk of the halo and thick disc population. Figure~\ref{fig:feh_kinematics} shows the distribution of the MSTO sample in the $L_z$–[Fe/H] and $V_{\phi}$–[Fe/H] planes, colour-coded by orbital eccentricity. It exhibits a low orbital eccentricity ($e \simeq 0.12$), a high azimuthal velocity ($V_{\phi} \simeq 220~\mathrm{km~s^{-1}}$), and a large positive angular momentum ($L_z \simeq 1755~\mathrm{km~kpc~s^{-1}}$), placing it firmly within the locus of dynamically cold, prograde orbits. Its guiding radius, $R_{\rm guid} \simeq 7.19\,\mathrm{kpc}$, lies close to the mean of the sample distribution ($\langle R_{\rm guid} \rangle \simeq 7.35\,\mathrm{kpc}$) and within its main peak, indicating an orbit with a guiding centre close to the solar neighbourhood. Its vertical orbital structure further supports this interpretation: the star reaches a maximum height above the Galactic plane of $Z_{\max} \simeq 0.28~\mathrm{kpc}$ and has a low vertical action ($J_z \simeq 2.7~\mathrm{km~kpc~s^{-1}}$), indicating that it remains tightly confined to the Galactic plane. This conclusion is reinforced by the Toomre diagram shown in Fig.~\ref{fig:toomre}, where the sample is colour-coded by metallicity. The candidate lies well within the low-velocity region associated with dynamically cold orbits, yet exhibits a high metallicity. This combination confirms the presence of metal-poor stars on thin disc-like orbits.

The orbital properties differ significantly from those typically observed in halo stars, which generally exhibit little net rotation, large velocity dispersions, highly eccentric orbits, and large vertical excursions from the Galactic plane \citep{Battaglia2005MNRAS.364..433B, Carollo2007Natur.450.1020C}. Adopting a canonical halo azimuthal velocity dispersion of \mbox{$\sigma_{\phi}\sim120\,\mathrm{km\,s^{-1}}$} \citep{Battaglia2005MNRAS.364..433B}, the observed rotational velocity corresponds to a $\sim1.8\sigma$ deviation from the halo mean. Under the assumption of a Gaussian distribution, the probability of finding a halo star with \mbox{$V_{\phi}\geq220\,\mathrm{km\,s^{-1}}$} is of the order of only a few percent ($\sim3\%$). However, when the additional orbital constraints are taken into account, namely the very small eccentricity and the low value of $Z_{\max}$, the likelihood of a halo origin decreases substantially. Halo stars typically reach several kiloparsecs above the Galactic plane and exhibit significantly hotter orbital distributions. Therefore, considering the full orbital configuration, we estimate that the probability of the star belonging to the halo population is likely below the percent level. The orbital properties are instead more naturally explained within the framework of the metal weak thick disc population, namely stars with halo-like metallicities but disc-like kinematics \citep{Morrison1990AJ....100.1191M, Chiba2000, Ruchti2010}. 

\begin{figure}
    \centering
    \includegraphics[width=\columnwidth]{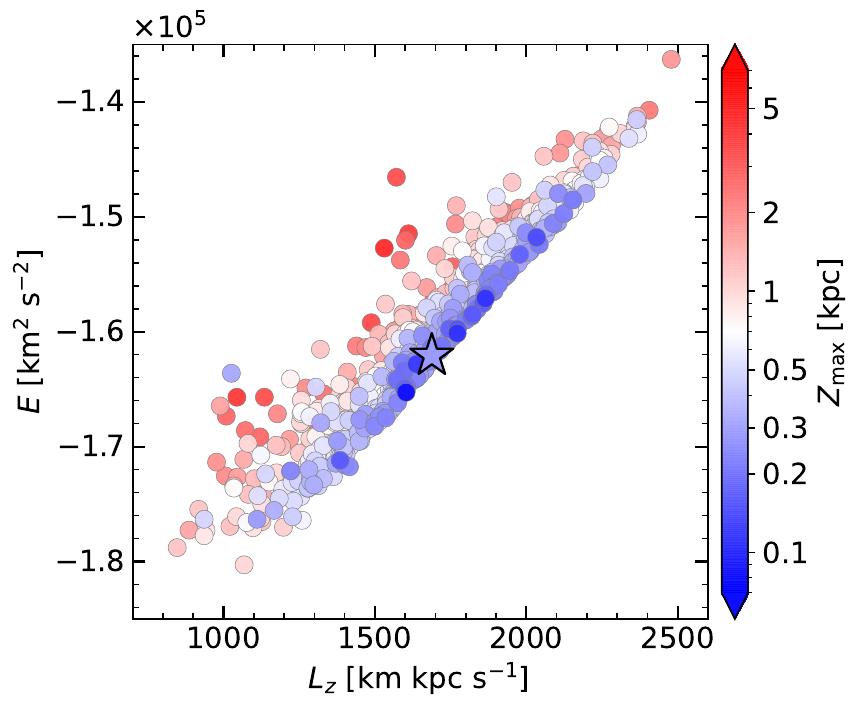}
    \caption{Lindblad diagram for the MSTO sample. The colour scale represents the maximum vertical excursion, $Z_{\max}$, highlighting the separation between stars confined to the Galactic plane and those on more vertically extended orbits. The candidate star is highlighted with a star symbol.}
    \label{fig:lindblad}
\end{figure}

A complementary view is provided by the Lindblad diagram (Fig.~\ref{fig:lindblad}), which shows the distribution of orbital energy as a function of angular momentum. The candidate lies along the main sequence defined by the bulk of the sample in the $E$--$L_z$ plane, characteristic of dynamically cold, rotationally supported orbits, and is clearly distinct from stars with lower angular momentum and higher energies typically associated with halo-like kinematics. This combination of low metallicity and dynamically cold, thin disc-like kinematics is uncommon and motivates a more detailed analysis of the nature of this object.

We conclude this section by noting that we also identified 17400785$-$6830303 as a metal-poor star with thin-disc-like orbital properties. However, this object is flagged in the Gaia-ESO recommended file as a spectroscopic binary candidate and also presents a technical convergence flag. This is relevant in light of the large number of spectroscopic-survey targets showing radial-velocity variability likely associated with their spectroscopic-binary nature \citep{VanDerSwaelmen2025}. In addition, its Gaia DR3 and Gaia-ESO radial velocities differ substantially, with $\mathrm{RV}_{\rm Gaia}=7.78$ km s$^{-1}$ and $\mathrm{RV}_{\rm GES}=45.11$ km s$^{-1}$, and Gaia DR3 flags the source as photometrically variable \citep{GaiaCollaboration2023}. We therefore do not consider this object as a robust candidate in the present analysis, despite its apparently cold prograde orbit. A similar kinematic search in the complete GES UVES field-star sample also revealed three additional metal-poor stars outside the MSTO selection with thin-disc-like orbital properties: 02195542$-$0446423, 08000517$-$0034337, and 18135851$-$4226346. These objects provide useful additional context and their evolutionary status is discussed in the following section. We note, however, that 18135851$-$4226346 is reported as a confirmed spectroscopic binary by \citet{Merle2017}, highlighting the need for caution when interpreting thin-disc-like orbital solutions for objects affected by binarity or possible radial-velocity variability.

\subsection{Age estimate and evolutionary status}

The age determination indicates that the target star is compatible with a very old stellar population. 
The posterior age distribution is broad and asymmetric, with a most probable value of $\mathrm{age}_{\rm mp}=10.2~\mathrm{Gyr}$. 
The corresponding 1$\sigma$-like percentiles extend from $\sim 4.4$ to $\sim 13.4~\mathrm{Gyr}$. We report the most probable age because the broad, asymmetric posterior makes the mode a more representative maximum-probability solution than the mean or median. Therefore, its qualitative properties are consistent with an old stellar population. 
When considered together with the star's dynamically cold and prograde orbit, this uncertain but old age estimate strengthens its interpretation as a candidate member of the metal-poor tail of the old thin disc. As shown in Fig.~\ref{fig:thin_disc_age_distribution}, the candidate lies at the old end of the age distribution of thin-disc stars selected according to kinematic and chemical criteria in our sample. The broad uncertainty of the age estimate is further illustrated in Fig.~\ref{fig:candidate_age_posterior_summary}, where we show a summary of the candidate's age posterior.

Among the three additional metal-poor stars identified in the complete GES UVES field-star sample with thin-disc-like orbital properties, only 18135851$-$4226346 has an age estimate compatible with an old thin-disc population. However, as discussed in subsection \ref{subsec:kin}, this object is a confirmed spectroscopic binary, while the other two objects do not show old age solutions. Therefore, none of these three additional stars is retained as a robust old metal-poor thin-disc candidate.

Finally, we also checked the Gaia DR3 astrometric quality indicators \citep{GaiaCollaboration2023}. The candidate has $\mathrm{RUWE}=1.366$, slightly elevated but below the commonly used $\mathrm{RUWE}\sim1.4$ threshold. The source also shows a significant astrometric excess noise 
($\epsilon_i=0.115$ mas, $\mathrm{D}=7.65$), although it is not flagged as duplicated 
($\texttt{duplicated\_source}=0$), has no multi-peak image detections 
($\texttt{ipd\_frac\_multi\_peak}=0$), and is not listed as a Gaia DR3 non-single star 
($\texttt{non\_single\_star}=0$). These indicators do not provide conclusive evidence for binarity, but unresolved multiplicity cannot be completely excluded.

\begin{table} 
\centering 
\caption{Stellar and orbital parameters of the candidate star.} 
\label{tab:target} \renewcommand{\arraystretch}{1.25} 
\begin{threeparttable}
\begin{tabular}{lc} 
\hline 
Parameter & Value \\ 
\hline 
CNAME & 15183399$-$0721310 \\ $T_{\rm eff}$ [K] & 5650$\pm$46 \\ $\log g$ & $4.11\pm0.07^{a}$ \\ $\mathrm{[Fe/H]}$ & $-1.38$$\pm$0.06 \\  $V_{\phi}$ [km s$^{-1}$] & $220.49^{+0.55}_{-0.26}$\\ $L_z$ [km\,kpc\,s$^{-1}$] & $1755.32^{+4.54}_{-2.74}$ \\ $E$ [$10^5$ km$^2$ s$^{-2}$] & $-1.6209^{+0.0015}_{-0.0011}$ \\ $R_{\rm guid}$ [kpc] & $7.191 \pm 0.016$ \\ $e$ & $0.1231^{+0.0040}_{-0.0037}$ \\ $Z_{\max}$ [kpc] & $0.2820^{+0.0067}_{-0.0075}$ \\ 
$J_z$ [km\,kpc\,s$^{-1}$] & $2.71^{+0.14}_{-0.16}$ \\
$\mathrm{age}_{\rm mp}$ [Gyr] & $10.2^{+3.2}_{-5.8}$ \\ 
$[\mathrm{Mg/Fe}]$ & $0.58\pm 0.11$ \\ $[\mathrm{Y/Mg}]$ & $-0.58\pm 0.10$ \\ $[\mathrm{Al/Fe}]$ & $0.27\pm 0.15$ \\ $[\mathrm{Mg/Mn}]$ & $0.65\pm 0.10$ \\ $A(\mathrm{Li})$ & $1.68\pm0.05$\\
$E(B-V)$ & $0.082^{b}$, $0.070^{c}$, $0.080^{d}$\\
\hline 
\end{tabular} 
\begin{tablenotes}[flushleft]
\footnotesize
\item \hspace{0.4cm}
Notes. $^{(a)}$ Spectroscopic value. The value inferred from the best-fitting isochrone is 
$\log g_{\rm iso}=4.57^{+0.06}_{-0.02}$.
$^{(b)}$ Estimated from \cite{sfd1998ApJ...500..525S}. 
$^{(c)}$ Scaled by 0.86 \citep{Schlafly2011ApJ...737..103S}. 
$^{(d)}$ Scaled to Bayestar \citep{green2018MNRAS.478..651G} according to the relation shown in Fig.~\ref{fig:ebv}.
\end{tablenotes}
\end{threeparttable}
\end{table}

\section{Chemical properties of the candidates}
\label{sec:chemistry}

Chemical abundances provide important constraints on the formation environment and enrichment history of stellar populations. In the following, the $\alpha$-element abundances of the candidates stars are examined first, followed by an analysis of the detailed chemical pattern and lithium abundance.

\subsection{$\alpha$-element abundances, the [Y/Mg]--age, and the [Al/Fe]--[Mg/Mn] plane}
\label{sec:YMG}
\begin{figure}
    \centering
    \includegraphics[width=\columnwidth]{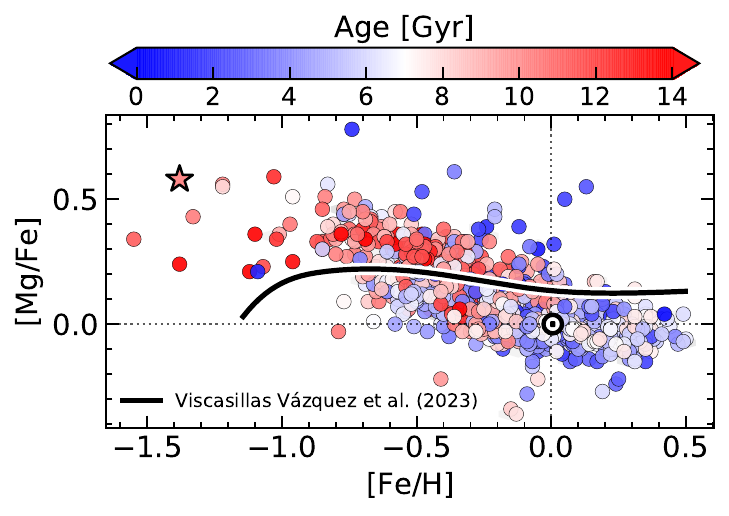}
    \caption{Distribution of the MSTO sample in the $[\mathrm{Mg/Fe}]$--$\mathrm{[Fe/H]}$ plane, colour-coded by stellar age. The star symbol corresponds to the candidate discussed in this work. The solid black curve marks the SVM-based classification boundary separating the thin-disc sequence from the thick-disc/halo locus taken from \citet{Viscasillas2023}}
    \label{fig:mgfe_age}
\end{figure}

We use Mg as a representative $\alpha$-element to investigate the chemical enrichment history of the candidate and of the comparison sample. Figure~\ref{fig:mgfe_age} shows the distribution of the MSTO sample in the $[\mathrm{Mg/Fe}]$--$[\mathrm{Fe/H}]$ plane, colour-coded by stellar age. The well-known $\alpha$-enhanced sequence is clearly visible. Using the Gaia-ESO Survey solar reference abundance, $A(\mathrm{Mg})_\odot = 7.51$, we derive $[\mathrm{Mg/H}] = -0.80$, corresponding to $[\mathrm{Mg/Fe}] = +0.58$ for the candidate. The candidate star therefore shows enhanced magnesium abundance relative to iron and lie within, or close to, the $\alpha$-enhanced sequence, overlapping with the oldest populations in the sample. The [$\alpha$/Fe] value computed from the other $\alpha$ elements, Si, Ca, and Ti, is 0.48, confirming that the star is $\alpha$ enhanced. This confirms that the candidate belongs to the old $\alpha$-enhanced population despite its thin-disc-like kinematics. In addition, these abundance ratios are in very good agreement with the independent values reported for the same star in GALAH DR4 \citep{Buder2025}, namely $[\mathrm{Fe/H}] = -1.47$, $[\mathrm{Mg/Fe}] = 0.59$, $[\mathrm{Si/Fe}] = 0.61$, $[\mathrm{Ca/Fe}] = 0.28$, and $[\mathrm{Ti/Fe}] = 0.31$. This is consistent with early and rapid chemical enrichment prior to the significant contribution of Type Ia supernovae. 

\begin{figure}
    \centering
    \includegraphics[width=\columnwidth]{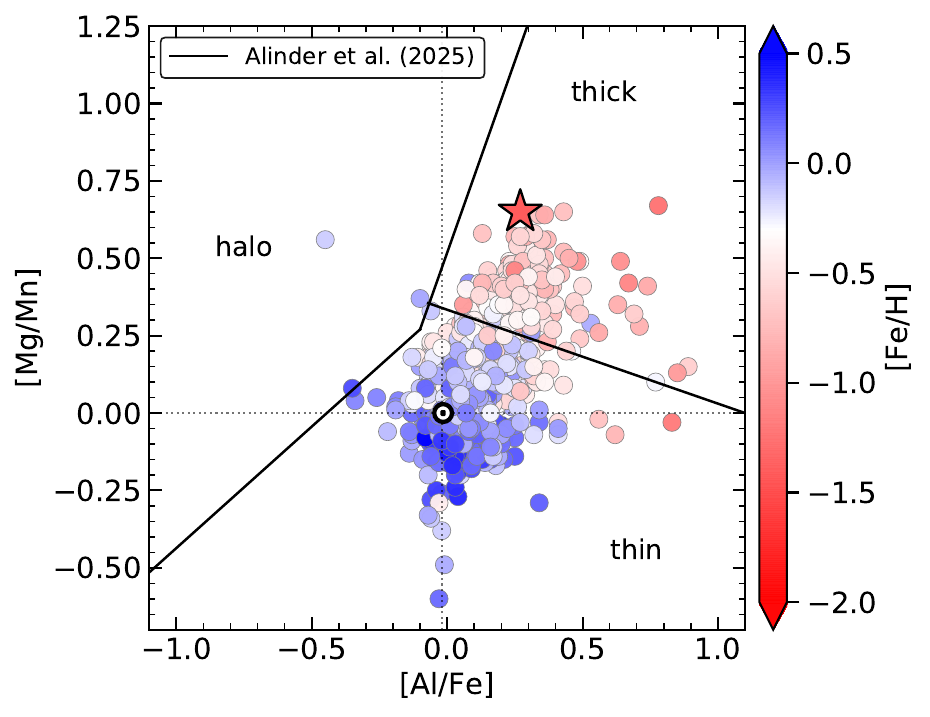}
    \caption{Distribution of the MSTO sample in the $[\mathrm{Al/Fe}]$--$[\mathrm{Mg/Mn}]$ plane, colour-coded by metallicity. The black solid lines indicate the empirical separation between Galactic components proposed by \citet{Alinder2026}. The candidate is marked with a star symbol.}
    \label{fig:alfe_mgmn}
\end{figure}

As an additional diagnostic, Figure~\ref{fig:alfe_mgmn} shows the position of the candidate in the $[\mathrm{Al/Fe}]$--$[\mathrm{Mg/Mn}]$ plane, colour-coded by metallicity. This plane has been shown to be particularly useful for separating Galactic populations and tracing different enrichment
histories \citep[e.g.][]{Hawkins2015,Das2020,Horta2021,Vasini2024,Alinder2026}. In the figure, we also show the chemical selection regions defined by \citet{Alinder2026}, which separate the regions mainly occupied by thin-disc,
thick-disc, and halo stars. This diagnostic is useful because $[\mathrm{Mg/Mn}]$ traces the relative contribution of core-collapse and Type Ia supernovae, while $[\mathrm{Al/Fe}]$ is sensitive to enrichment by massive stars and to the intensity of star formation \citep[e.g.][]{Kobayashi2006,Mishenina2015}. In this diagram, the candidate lies in the region associated with the thick disc, close to the halo boundary, and clearly separated from the thin-disc locus. Its relatively high $[\mathrm{Mg/Mn}]$ and enhanced $[\mathrm{Al/Fe}]$, together with its low metallicity, therefore support the view that the star formed from gas enriched predominantly by massive stars on short timescales, rather than following the chemical evolution typical of the canonical thin disc.

\begin{figure}
    \centering
    \includegraphics[width=\columnwidth]{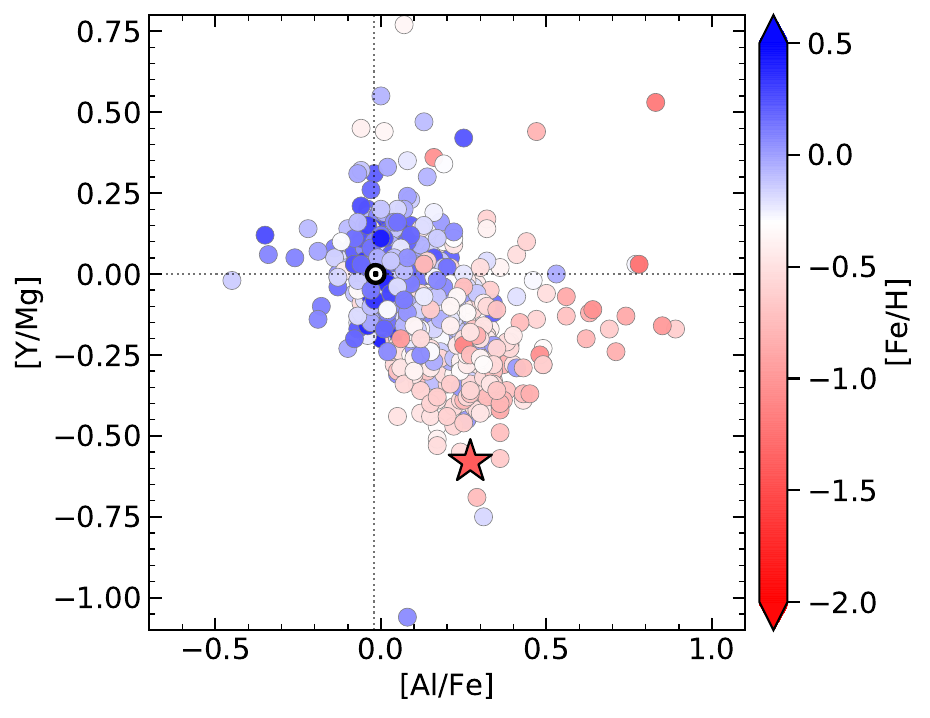}
    \caption{Distribution of the sample in the $[\mathrm{Al/Fe}]$--$[\mathrm{Y/Mg}]$
    plane, colour-coded by $[\mathrm{Fe/H}]$.}
    \label{fig:alfe_ymg}
\end{figure}

The abundance pattern in the $[\mathrm{Al/Fe}]$--$[\mathrm{Y/Mg}]$ plane
provides an additional indication of an early formation epoch. The candidate
shows enhanced $[\mathrm{Al/Fe}]$ together with a very low $[\mathrm{Y/Mg}] = -0.58$ (Fig.~\ref{fig:alfe_ymg}). The former points to efficient enrichment by massive stars and core-collapse supernovae, while the
latter indicates that the gas from which the star formed had received only a
limited contribution from the delayed production of Y by low- and intermediate-mass AGB stars \citep[e.g.][]{Travaglio2004,Bisterzo2014}. Low
$[\mathrm{Y/Mg}]$ ratios are generally associated with old stellar populations,
reflecting the different enrichment timescales of Mg, mainly produced by massive
stars, and Y, largely contributed by the slow neutron-capture process in AGB
stars \citep[e.g.][]{Nissen2015,Feltzing2017,DelgadoMena2019,
Viscasillas2022, Molero2025}. Its $[\mathrm{Y/Mg}]$ ratio lies below the values typically observed for thin-disc stars in the relations of
\citet{Tautvaisiene2021, Viscasillas2025,Pakstiene2026, Mikolaitis2026}, reinforcing the idea that, despite its thin-disc-like orbit, the star is chemically associated with an old
population formed during the early phases of Galactic disc assembly.

\subsection{Detailed chemical pattern}
We analyse the detailed chemical pattern of the candidate using abundance ratios [El/Fe] for elements tracing different nucleosynthetic channels, following \citet{Magrini2023}. The abundances are normalised to the Gaia-ESO Survey solar reference scale. The candidate is then compared with four reference stars selected according to their kinematic properties.
 
We first include a representative thin-disc star selected from the subsample satisfying both the kinematic and chemical thin-disc criteria.  Within this subsample, we chose the object closest to the median metallicity and the mean most-probable age of the thin-disc population. The thin-disc subsample has a mean metallicity of $[\mathrm{Fe/H}] = -0.025$ and $\langle \mathrm{age}_{\rm mp} \rangle = 5.09$ Gyr, with a dispersion of $2.37$ Gyr. The selected thin-disc comparison star, CNAME 17390751$-$5951521, has $[\mathrm{Fe/H}]=-0.07$ and $\mathrm{age}_{\rm mp}=5.8^{+1.6}_{-1.7}$ Gyr.

The initial selection of the thick-disc and halo comparison stars was guided by proximity to the candidate in the age--metallicity plane, using the normalised Euclidean distance

\begin{equation}
d =
\sqrt{
\left(\frac{\mathrm{age}-\mathrm{age}_{\rm ref}}{1~\mathrm{Gyr}}\right)^2 +
\left(\frac{[\mathrm{Fe/H}]-[\mathrm{Fe/H}]_{\rm ref}}{0.10~\mathrm{dex}}\right)^2
}
\label{eq:age_feh_distance}
\end{equation}

For the thin-disc comparison, the reference values were the mean most-probable age and mean metallicity of the subsample selected according to both kinematic and chemical thin-disc criteria. For the thick-disc and halo comparisons, the reference values were those of the candidate. The final comparison stars were then required to be representative of their respective populations in both kinematics and chemistry. The selected thick-disc star is CNAME 01393038$-$5400563, with $\mathrm{age}_{\rm mp}=10.6^{+1.2}_{-1.5}$ Gyr, $[\mathrm{Fe/H}]=-0.73$, and $V_{\rm tot}=176.25$ km s$^{-1}$. For the halo comparison, the number of suitable stars was very limited. 
Only a few objects in our sample have halo-like velocities ($V_{\rm tot} > 180$ km s$^{-1}$) and metallicities below $[\mathrm{Fe/H}]<-1.0$. 
We first excluded chemically peculiar objects as the metallicity-matched star CNAME 03373965$-$2721214, which has $[\mathrm{Fe/H}]=-1.38$ as the candidate, but is flagged in SIMBAD as chemically peculiar  \citep[see e.g.][]{Fu2018, Magrini2021}. Among the remaining non-peculiar halo-kinematic stars, CNAME 04273549$-$3505248 was selected as the best compromise between metallicity and abundance-pattern similarity to the candidate. This halo comparison star has $\mathrm{age}_{\rm mp}=10.4^{+0.9}_{-0.7}$ Gyr, $[\mathrm{Fe/H}]=-1.22$, and $V_{\rm tot}=290.88$ km s$^{-1}$.

The resulting abundance patterns are shown in Fig.~\ref{fig:xfe_pattern} in Appendix. The representative thin-disc star displays a clearly different pattern, as expected. 
By contrast, the candidate is broadly more similar to the old, metal-poor thick-disc and halo comparison stars, particularly in the $\alpha$-elements and several iron-peak species. The largest differences are found in some iron-peak and neutron-capture elements, particularly among the heavy elements at the end of the sequence. The representative thin-disc star shows the most distinct pattern, lying systematically below the candidate for several neutron-capture species. By contrast, the thick-disc and halo comparison stars are generally closer to the candidate for the light and $\alpha$-elements. In particular, the comparison halo star provides the closest match to the candidate in metallicity and chemistry, with very similar $[\mathrm{Mg/Fe}]$ and $[\mathrm{Mg/Mn}]$ ratios: 0.56 and 0.67, compared to 0.58 and 0.65 for the candidate. However, this chemical similarity contrasts with their markedly different kinematics: the comparison star has halo-like orbital properties ($V_{\phi}=-35.05\,\mathrm{km\,s^{-1}}$, $L_z=-311.16\,\mathrm{kpc\,km\,s^{-1}}$, $e=0.86$, $Z_{\max}=1.26\,\mathrm{kpc}$), whereas the candidate displays thin-disc-like kinematics (see Table~\ref{tab:target}). 

On the whole, the abundance pattern of the candidate is more consistent with old, metal-poor populations than with the typical thin disc. This supports the interpretation that, despite its thin-disc-like kinematics, the star formed from gas enriched under conditions characteristic of the early Galaxy, and may trace the metal-poor tail of the old thin disc or an early phase of the Galactic disc formation.

\subsection{Lithium abundance}

\begin{figure}
    \centering
    \includegraphics[width=\columnwidth]{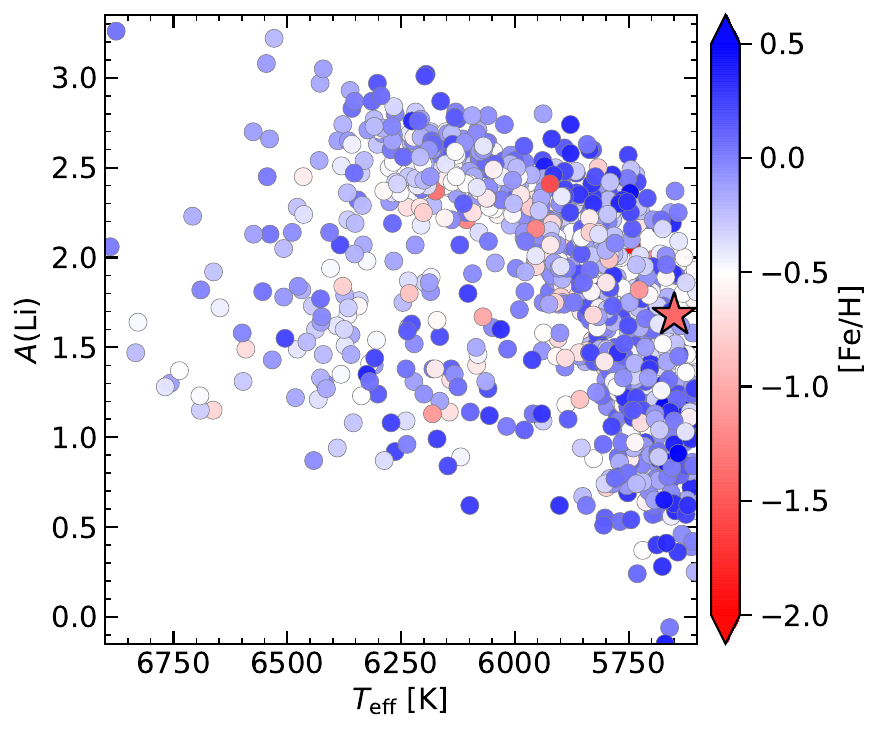}
    \caption{Lithium abundance as a function of effective temperature for the MSTO sample, colour-coded by metallicity. The star symbol corresponds to the candidate discussed in this work.}
    \label{fig:li_feh_teff}
\end{figure}

To further characterise the nature of the candidate, we analyse its lithium abundance in the context of the MSTO sample. Figure~\ref{fig:li_feh_teff} shows the lithium abundance as a function of effective temperature, colour coded by metallicity. The Spite plateau is primarily traced by the warmest stars and most metal poor stars in the sample, while cooler stars exhibit progressively lower lithium abundances due to depletion. The candidate is located at $\mathrm{[Fe/H]} = -1.38$ and has an effective temperature of $T_{\rm eff} = 5650\,\mathrm{K}$; its measured lithium abundance, $A(\mathrm{Li}) = 1.68 \pm 0.05$, lies below the canonical {\it Spite plateau} value \citep[e.g.][and references therein]{Yang2026}. This is consistent with previous findings in the literature \citep[e.g.]{cp2005} for metal-poor halo stars with similar temperatures; interestingly, the Li value for the candidate is consistent with recent model predictions for Li depletion in metal poor stars \citep{Yang2026} and an initial Li corresponding A(Li)$=2.72$ as predicted by Big Bang Nucleosynthesis. 

\section{Discussion}
\label{sec:discussion}

The Gaia-ESO candidate is a metal-poor star exhibiting dynamically cold, thin disc-like kinematics, a combination that is uncommon and particularly informative for understanding the early formation of the Galactic disc. With $\mathrm{[Fe/H]} = -1.38$, the star lies in a metallicity regime where halo populations are expected to dominate in the solar neighbourhood (e.g. \citealt{Nissen2010, Haywood2013, Helmi2018, Belokurov2018}). However, its orbital properties—including low eccentricity, small vertical excursion, high prograde velocity, and large angular momentum—are fully consistent with a dynamically cold, rotationally supported orbit \citep[e.g.][]{Bensby2003, Bensby2014}. Such properties place the star within the population of metal-poor, high-angular-momentum stars identified in recent large spectroscopic surveys (e.g. \citealt{FernandezAlvar2021, Hu2025}). These studies reveal a sequence of stars with thin-disc-like kinematics extending to low metallicities, although their origin remains debated. In particular, it is still unclear whether these objects represent the metal-poor extension of the thin disc or the high-angular-momentum tail of dynamically hotter populations. In this context, the possibility of a metal-weak thick-disc origin cannot be excluded 
\citep[e.g.][]{Zhang2024}, as the chemically selected Mg-enhanced population in our sample also includes a high-$V_{\phi}$ tail comparable to the candidate. This is illustrated in Fig.~\ref{fig:mg_enhanced_kinematics} in Appendix~\ref{app:mg_enhanced_kinematics}, which shows the $V_{\phi}$, $V_{\rm tot}$, and $Z_{\max}$ distributions of this chemically selected sample after removing young $\alpha$-rich stars. However, considering its overall orbital properties, the candidate lies among the dynamically colder objects of this population. This ambiguity is consistent with both observational studies of metal-poor stars in the solar neighbourhood, where chemical and orbital classifications may differ \citep[e.g.][]{Sahin2020}, and numerical models showing that present-day kinematics do not necessarily preserve a direct record of the original epoch of disc formation after merger-driven perturbations \citep[e.g.][]{Orkney2026}.

Recent work has provided a time-resolved view of Milky Way assembly, revealing a transition from an early $\alpha$-enhanced phase to the later emergence of a dynamically cold disc, with a characteristic separation at $\sim$8 Gyr \citep{Xiang2022}. The candidate, whose age posterior favours an old solution with a most probable value of $10.2^{+3.2}_{-5.8}$ Gyr, low metallicity and enhanced $\alpha$-abundances, is compatible with this early phase, yet already exhibits disc-like kinematics, suggesting that rotationally supported orbits were established at very early epochs. In the context of recent work on the early Milky Way, this object can be interpreted within the transition between the kinematically hot, metal-poor in-situ population (often referred to as \emph{Aurora}) and the onset of disc formation \citep{Belokurov2022}. Stars in the Aurora component are characterised by dynamically hot orbits and modest net rotation, while a rapid increase in rotational support (the so-called ``spin-up'') occurs over a relatively short timescale as the Galactic disc forms. The kinematics of the candidate, combined with its metallicity, suggest that it may belong to this early disc phase, tracing the emergence of dynamically cold orbits shortly after the spin-up of the Galaxy.

The chemical properties of the star provide a key complementary constraint. Its low metallicity $([\mathrm{Fe/H}]=-1.38)$, strong $\alpha$-enhancement ($[\mathrm{Mg/Fe}] = +0.58$; [$\mathrm{\alpha/Fe}]=+0.48$) and old age estimate of $10.2^{+3.2}_{-5.8}$ Gyr point to formation during a phase of rapid chemical enrichment dominated by core-collapse supernovae, typical of early in-situ populations. This picture is further supported by its very low $[\mathrm{Y/Mg}]$ ratio 
($[\mathrm{Y/Mg}] = -0.58$), which is consistent with formation before the delayed contribution of s-process material from low- and intermediate-mass AGB stars became significant. These properties are not consistent with the canonical thin disc, but overlap with those of the metal-weak thick disc and other chemically old Galactic components (e.g. \citealt{Chiba2000,Reddy2006}). This interpretation is further supported by its detailed abundance pattern, which is broadly consistent with that of halo stars across a wide range of elements tracing different nucleosynthetic channels. In this framework, the candidate does not appear to be a member of the present-day thin disc, but rather traces an early stage in the assembly of the Galactic disc. The lithium abundance provides an additional constraint on its nature. 

\section{Summary and conclusions}
\label{sec:summary}
We have identified a metal-poor star ($\mathrm{[Fe/H]} = -1.38$) with dynamically cold, thin disc-like kinematics, including low eccentricity, small vertical excursion, and high prograde rotational velocity. These properties place the star within the population of metal-poor objects with disc-like rotational support recently identified in large surveys.

Despite its thin-disc-like kinematics, the star exhibits $\alpha$-enhanced chemical abundances and an age estimate consistent with an old stellar population, with a most probable value of $\mathrm{age}_{\rm mp}=10.2^{+3.2}_{-5.8}$ Gyr, indicating formation during the early phases of Galactic evolution. Its detailed chemical pattern across multiple elements is broadly consistent with that of halo stars, reinforcing an origin in gas enriched predominantly by core-collapse supernovae. This combination suggests that the star does not belong to the canonical thin disc. Instead, the candidate is more naturally interpreted as part of a transitional population tracing the onset of disc formation, likely associated with the rapid spin-up phase of the Milky Way. Its properties indicate that dynamically cold, disc-like orbits were already in place at early times. 
The lithium abundance of the star, $A(\mathrm{Li}) = 1.68$, is also consistent with the halo, rather than thin disc population.

This object provides direct observational evidence that metal-poor, $\alpha$-enhanced stars can exhibit thin disc-like kinematics, offering new constraints on the timing and dynamical settling of the Galactic disc.

\begin{acknowledgements}
We thank the anonymous referee for their constructive comments and suggestions, which helped improve the clarity and interpretation of this work. CVV acknowledges support from the Erasmus+ Staff, which enabled a scientific visit to the Arcetri Astrophysical Observatory -- INAF, within the framework of KA1 Learning Mobility of Individuals. 
R.E.G., L.B. and L.M. acknowledge support from INAF through the Large Grants EPOCH and WST, funding for the WEAVE project, the Mini-Grants Checs (1.05.23.04.02), and financial support under the National Recovery and Resilience Plan (PNRR), Mission 4, Component 2, Investment 1.1, Call for tender No. 104 published on 2 February 2022 by the Italian Ministry of University and Research (MUR), funded by the European Union – NextGenerationEU, through the Project ‘Cosmic POT’ (Grant Assignment Decree No. 2022X4TM3H, MUR). Use was made of the Simbad database, operated at the CDS, Strasbourg, France, and of NASA’s Astrophysics Data System Bibliographic Services. 
This work is based on data products from observations made with ESO Telescopes at the La Silla Paranal Observatory under programmes 188.B-3002, 193.B-0936, and 197.B-1074.
\end{acknowledgements}

\bibliographystyle{aa} 
\bibliography{bibliography}

@ARTICLE{Morrison1990AJ....100.1191M,
       author = {{Morrison}, Heather L. and {Flynn}, Chris and {Freeman}, K.~C.},
        title = "{Where Does the Disk Stop and the Halo Begin - Kinematics in a Rotation Field}",
      journal = {\aj},
         year = 1990,
        month = oct,
       volume = {100},
        pages = {1191},
          doi = {10.1086/115587},
       adsurl = {https://ui.adsabs.harvard.edu/abs/1990AJ....100.1191M}
}

@ARTICLE{Battaglia2005MNRAS.364..433B,
       author = {{Battaglia}, Giuseppina and {Helmi}, Amina and {Morrison}, Heather and {Harding}, Paul and {Olszewski}, Edward W. and {Mateo}, Mario and {Freeman}, Kenneth C. and {Norris}, John and {Shectman}, Stephen A.},
        title = "{The radial velocity dispersion profile of the Galactic halo: constraining the density profile of the dark halo of the Milky Way}",
      journal = {\mnras},
         year = 2005,
        month = dec,
       volume = {364},
       number = {2},
        pages = {433-442},
          doi = {10.1111/j.1365-2966.2005.09367.x},
archivePrefix = {arXiv},
       eprint = {astro-ph/0506102},
 primaryClass = {astro-ph},
       adsurl = {https://ui.adsabs.harvard.edu/abs/2005MNRAS.364..433B}
}

@ARTICLE{Carollo2007Natur.450.1020C,
       author = {{Carollo}, Daniela and {Beers}, Timothy C. and {Lee}, Young Sun and {Chiba}, Masashi and {Norris}, John E. and {Wilhelm}, Ronald and {Sivarani}, Thirupathi and {Marsteller}, Brian and {Munn}, Jeffrey A. and {Bailer-Jones}, Coryn A.~L. and {Fiorentin}, Paola Re and {York}, Donald G.},
        title = "{Two stellar components in the halo of the Milky Way}",
      journal = {\nat},
         year = 2007,
        month = dec,
       volume = {450},
       number = {7172},
        pages = {1020-1025},
          doi = {10.1038/nature06460},
archivePrefix = {arXiv},
       eprint = {0706.3005},
 primaryClass = {astro-ph},
       adsurl = {https://ui.adsabs.harvard.edu/abs/2007Natur.450.1020C}
}

@ARTICLE{Bellazzini2024,
       author = {{Bellazzini}, M. and {Massari}, D. and {Ceccarelli}, E. and {Mucciarelli}, A. and {Bragaglia}, A. and {Riello}, M. and {De Angeli}, F. and {Montegriffo}, P.},
        title = "{Metal-poor stars with disc-like orbits. Possible traces of the Galactic disc at very early epochs}",
      journal = {\aap},
         year = 2024,
        month = mar,
       volume = {683},
          eid = {A136},
        pages = {A136},
          doi = {10.1051/0004-6361/202348106},
archivePrefix = {arXiv},
       eprint = {2312.02356},
 primaryClass = {astro-ph.GA},
       adsurl = {https://ui.adsabs.harvard.edu/abs/2024A&A...683A.136B}
}

@ARTICLE{Hu2025,
       author = {{Hu}, Guozhen and {Shao}, Zhengyi and {G{\"u}gercino{\v{g}}lu}, Erbil and {Cui}, Wenyuan},
        title = "{Investigating the Lower Limit of Metallicity for the Galactic Thin Disk}",
      journal = {\apj},
         year = 2025,
        month = jan,
       volume = {979},
       number = {1},
          eid = {59},
        pages = {59},
          doi = {10.3847/1538-4357/ad9c38},
archivePrefix = {arXiv},
       eprint = {2412.06187},
 primaryClass = {astro-ph.GA},
       adsurl = {https://ui.adsabs.harvard.edu/abs/2025ApJ...979...59H}
}

@ARTICLE{Yang2026,
       author = {{Yang}, Wuming and {Dou}, Shuya and {Meng}, Xiangcun and {Wu}, Yaqian and {Bi}, Shaolan},
        title = "{Using Lithium and Beryllium to Study the Structure and Evolution of Rotating Stars: The Spite Plateau of Halo Stars}",
      journal = {\apj},
         year = 2026,
        month = apr,
       volume = {1000},
       number = {2},
          eid = {271},
        pages = {271},
          doi = {10.3847/1538-4357/ae4d0f},
archivePrefix = {arXiv},
       eprint = {2602.22516},
 primaryClass = {astro-ph.SR},
       adsurl = {https://ui.adsabs.harvard.edu/abs/2026ApJ..1000..271Y}
}

@ARTICLE{cp2005,
       author = {{Charbonnel}, C. and {Primas}, F.},
        title = "{The lithium content of the Galactic Halo stars}",
      journal = {\aap},
         year = 2005,
        month = nov,
       volume = {442},
       number = {3},
        pages = {961-992},
          doi = {10.1051/0004-6361:20042491},
archivePrefix = {arXiv},
       eprint = {astro-ph/0505247},
 primaryClass = {astro-ph},
       adsurl = {https://ui.adsabs.harvard.edu/abs/2005A&A...442..961C}
}

@ARTICLE{Sestito2019,
       author = {{Sestito}, Federico and {Longeard}, Nicolas and {Martin}, Nicolas F. and {Starkenburg}, Else and {Fouesneau}, Morgan and {Gonz{\'a}lez Hern{\'a}ndez}, Jonay I. and {Arentsen}, Anke and {Ibata}, Rodrigo and {Aguado}, David S. and {Carlberg}, Raymond G. and {Jablonka}, Pascale and {Navarro}, Julio F. and {Tolstoy}, Eline and {Venn}, Kim A.},
        title = "{Tracing the formation of the Milky Way through ultra metal-poor stars}",
      journal = {\mnras},
         year = 2019,
        month = apr,
       volume = {484},
       number = {2},
        pages = {2166-2180},
          doi = {10.1093/mnras/stz043},
archivePrefix = {arXiv},
       eprint = {1811.03099},
 primaryClass = {astro-ph.GA},
       adsurl = {https://ui.adsabs.harvard.edu/abs/2019MNRAS.484.2166S}
}

@ARTICLE{Sestito2020,
       author = {{Sestito}, Federico and {Martin}, Nicolas F. and {Starkenburg}, Else and {Arentsen}, Anke and {Ibata}, Rodrigo A. and {Longeard}, Nicolas and {Kielty}, Collin and {Youakim}, Kristopher and {Venn}, Kim A. and {Aguado}, David S. and {Carlberg}, Raymond G. and {Gonz{\'a}lez Hern{\'a}ndez}, Jonay I. and {Hill}, Vanessa and {Jablonka}, Pascale and {Kordopatis}, Georges and {Malhan}, Khyati and {Navarro}, Julio F. and {S{\'a}nchez-Janssen}, Rub{\'e}n and {Thomas}, Guillame and {Tolstoy}, Eline and {Wilson}, Thomas G. and {Palicio}, Pedro A. and {Bialek}, Spencer and {Garcia-Dias}, Rafael and {Lucchesi}, Romain and {North}, Pierre and {Osorio}, Yeisson and {Patrick}, Lee R. and {Peralta de Arriba}, Luis},
        title = "{The Pristine survey - X. A large population of low-metallicity stars permeates the Galactic disc}",
      journal = {\mnras},
         year = 2020,
        month = sep,
       volume = {497},
       number = {1},
        pages = {L7-L12},
          doi = {10.1093/mnrasl/slaa022},
archivePrefix = {arXiv},
       eprint = {1911.08491},
 primaryClass = {astro-ph.GA},
       adsurl = {https://ui.adsabs.harvard.edu/abs/2020MNRAS.497L...7S}
}

@ARTICLE{Cordoni2021,
       author = {{Cordoni}, G. and {Da Costa}, G.~S. and {Yong}, D. and {Mackey}, A.~D. and {Marino}, A.~F. and {Monty}, S. and {Nordlander}, T. and {Norris}, J.~E. and {Asplund}, M. and {Bessell}, M.~S. and {Casey}, A.~R. and {Frebel}, A. and {Lind}, K. and {Murphy}, S.~J. and {Schmidt}, B.~P. and {Gao}, X.~D. and {Xylakis-Dornbusch}, T. and {Amarsi}, A.~M. and {Milone}, A.~P.},
        title = "{Exploring the Galaxy's halo and very metal-weak thick disc with SkyMapper and Gaia DR2}",
      journal = {\mnras},
         year = 2021,
        month = may,
       volume = {503},
       number = {2},
        pages = {2539-2561},
          doi = {10.1093/mnras/staa3417},
archivePrefix = {arXiv},
       eprint = {2011.01189},
 primaryClass = {astro-ph.GA},
       adsurl = {https://ui.adsabs.harvard.edu/abs/2021MNRAS.503.2539C}
}

@ARTICLE{Carter2021,
       author = {{Carter}, Courtney and {Conroy}, Charlie and {Zaritsky}, Dennis and {Ting}, Yuan-Sen and {Bonaca}, Ana and {Naidu}, Rohan P. and {Johnson}, Benjamin D. and {Cargile}, Phillip A. and {Caldwell}, Nelson and {Speagle}, Josh and {Han}, Jiwon Jesse},
        title = "{Ancient Very Metal-poor Stars Associated with the Galactic Disk in the H3 Survey}",
      journal = {\apj},
         year = 2021,
        month = feb,
       volume = {908},
       number = {2},
          eid = {208},
        pages = {208},
          doi = {10.3847/1538-4357/abcda4},
archivePrefix = {arXiv},
       eprint = {2012.00036},
 primaryClass = {astro-ph.GA},
       adsurl = {https://ui.adsabs.harvard.edu/abs/2021ApJ...908..208C}
}

@ARTICLE{Carollo2023,
       author = {{Carollo}, Daniela and {Christlieb}, Norbert and {Tissera}, Patricia B. and {Sillero}, Emanuel},
        title = "{Understanding the Early Stages of Galaxy Formation Using Very Metal-poor Stars from the Hamburg/ESO Survey}",
      journal = {\apj},
         year = 2023,
        month = apr,
       volume = {946},
       number = {2},
          eid = {99},
        pages = {99},
          doi = {10.3847/1538-4357/acac25},
archivePrefix = {arXiv},
       eprint = {2212.08294},
 primaryClass = {astro-ph.GA},
       adsurl = {https://ui.adsabs.harvard.edu/abs/2023ApJ...946...99C}
}

@ARTICLE{FernandezAlvar2021,
       author = {{Fern{\'a}ndez-Alvar}, Emma and {Kordopatis}, Georges and {Hill}, Vanessa and {Starkenburg}, Else and {Viswanathan}, Akshara and {Martin}, Nicolas F. and {Thomas}, Guillaume F. and {Navarro}, Julio F. and {Malhan}, Khyati and {Sestito}, Federico and {Gonz{\'a}lez Hern{\'a}ndez}, Jonay I. and {Carlberg}, Raymond G.},
        title = "{The Pristine survey XIII: uncovering the very metal-poor tail of the thin disc}",
      journal = {\mnras},
         year = 2021,
        month = nov,
       volume = {508},
       number = {1},
        pages = {1509-1525},
          doi = {10.1093/mnras/stab2617},
archivePrefix = {arXiv},
       eprint = {2106.03406},
 primaryClass = {astro-ph.GA},
       adsurl = {https://ui.adsabs.harvard.edu/abs/2021MNRAS.508.1509F}
}

@ARTICLE{FernandezAlvar2024,
       author = {{Fern{\'a}ndez-Alvar}, Emma and {Kordopatis}, Georges and {Hill}, Vanessa and {Battaglia}, Giuseppina and {Gallart}, Carme and {Gonz{\'a}lez Rivera de la Vernhe}, Isaure and {Thomas}, Guillaume and {Sestito}, Federico and {Ardern-Arentsen}, Anke and {Martin}, Nicolas and {Viswanathan}, Akshara and {Starkenburg}, Else},
        title = "{The metal-poor edge of the Milky Way's ``thin disc''}",
      journal = {\aap},
         year = 2024,
        month = may,
       volume = {685},
          eid = {A151},
        pages = {A151},
          doi = {10.1051/0004-6361/202348918},
archivePrefix = {arXiv},
       eprint = {2402.02943},
 primaryClass = {astro-ph.GA},
       adsurl = {https://ui.adsabs.harvard.edu/abs/2024A&A...685A.151F}
}

@ARTICLE{ArdernArentsen2024,
       author = {{Ardern-Arentsen}, Anke and {Monari}, Giacomo and {Queiroz}, Anna B.~A. and {Starkenburg}, Else and {Martin}, Nicolas F. and {Chiappini}, Cristina and {Aguado}, David S. and {Belokurov}, Vasily and {Carlberg}, Ray and {Monty}, Stephanie and {Myeong}, GyuChul and {Schultheis}, Mathias and {Sestito}, Federico and {Venn}, Kim A. and {Vitali}, Sara and {Yuan}, Zhen and {Zhang}, Hanyuan and {Buder}, Sven and {Lewis}, Geraint F. and {Oliver}, William H. and {Wan}, Zhen and {Zucker}, Daniel B.},
        title = "{The Pristine Inner Galaxy Survey - VIII. Characterizing the orbital properties of the ancient, very metal-poor inner Milky Way}",
      journal = {\mnras},
         year = 2024,
        month = may,
       volume = {530},
       number = {3},
        pages = {3391-3411},
          doi = {10.1093/mnras/stae1049},
archivePrefix = {arXiv},
       eprint = {2312.03847},
 primaryClass = {astro-ph.GA},
       adsurl = {https://ui.adsabs.harvard.edu/abs/2024MNRAS.530.3391A}
}

@ARTICLE{Nepal2024,
       author = {{Nepal}, S. and {Chiappini}, C. and {Queiroz}, A.~B. and {Guiglion}, G. and {Montalb{\'a}n}, J. and {Steinmetz}, M. and {Miglio}, A. and {Khalatyan}, A.},
        title = "{Discovery of the local counterpart of disc galaxies at z > 4: The oldest thin disc of the Milky Way using Gaia-RVS}",
      journal = {\aap},
         year = 2024,
        month = aug,
       volume = {688},
          eid = {A167},
        pages = {A167},
          doi = {10.1051/0004-6361/202449445},
archivePrefix = {arXiv},
       eprint = {2402.00561},
 primaryClass = {astro-ph.GA},
       adsurl = {https://ui.adsabs.harvard.edu/abs/2024A&A...688A.167N}
}

@ARTICLE{Neeleman2020,
       author = {{Neeleman}, Marcel and {Prochaska}, J. Xavier and {Kanekar}, Nissim and {Rafelski}, Marc},
        title = "{A cold, massive, rotating disk galaxy 1.5 billion years after the Big Bang}",
      journal = {\nat},
         year = 2020,
        month = may,
       volume = {581},
       number = {7808},
        pages = {269-272},
          doi = {10.1038/s41586-020-2276-y},
archivePrefix = {arXiv},
       eprint = {2005.09661},
 primaryClass = {astro-ph.GA},
       adsurl = {https://ui.adsabs.harvard.edu/abs/2020Natur.581..269N}
}

@ARTICLE{Rizzo2020,
       author = {{Rizzo}, F. and {Vegetti}, S. and {Powell}, D. and {Fraternali}, F. and {McKean}, J.~P. and {Stacey}, H.~R. and {White}, S.~D.~M.},
        title = "{A dynamically cold disk galaxy in the early Universe}",
      journal = {\nat},
         year = 2020,
        month = aug,
       volume = {584},
       number = {7820},
        pages = {201-204},
          doi = {10.1038/s41586-020-2572-6},
archivePrefix = {arXiv},
       eprint = {2009.01251},
 primaryClass = {astro-ph.GA},
       adsurl = {https://ui.adsabs.harvard.edu/abs/2020Natur.584..201R}
}

@ARTICLE{Ferreira2022,
       author = {{Ferreira}, Leonardo and {Adams}, Nathan and {Conselice}, Christopher J. and {Sazonova}, Elizaveta and {Austin}, Duncan and {Caruana}, Joseph and {Ferrari}, Fabricio and {Verma}, Aprajita and {Trussler}, James and {Broadhurst}, Tom and {Diego}, Jose and {Frye}, Brenda L. and {Pascale}, Massimo and {Wilkins}, Stephen M. and {Windhorst}, Rogier A. and {Zitrin}, Adi},
        title = "{Panic! at the Disks: First Rest-frame Optical Observations of Galaxy Structure at z > 3 with JWST in the SMACS 0723 Field}",
      journal = {\apjl},
         year = 2022,
        month = oct,
       volume = {938},
       number = {1},
          eid = {L2},
        pages = {L2},
          doi = {10.3847/2041-8213/ac947c},
archivePrefix = {arXiv},
       eprint = {2207.09428},
 primaryClass = {astro-ph.GA},
       adsurl = {https://ui.adsabs.harvard.edu/abs/2022ApJ...938L...2F}
}

@ARTICLE{Kartaltepe2023,
       author = {{Kartaltepe}, Jeyhan S. and {Rose}, Caitlin and {Vanderhoof}, Brittany N. and {McGrath}, Elizabeth J. and {Costantin}, Luca and {Cox}, Isabella G. and {Yung}, L.~Y. Aaron and {Kocevski}, Dale D. and {Wuyts}, Stijn and {Ferguson}, Henry C. and {Bagley}, Micaela B. and {Finkelstein}, Steven L. and {Amor{\'\i}n}, Ricardo O. and {Andrews}, Brett H. and {Arrabal Haro}, Pablo and {Backhaus}, Bren E. and {Behroozi}, Peter and {Bisigello}, Laura and {Calabr{\`o}}, Antonello and {Casey}, Caitlin M. and {Coogan}, Rosemary T. and {Cooper}, M.~C. and {Croton}, Darren and {de la Vega}, Alexander and {Dickinson}, Mark and {Fontana}, Adriano and {Franco}, Maximilien and {Grazian}, Andrea and {Grogin}, Norman A. and {Hathi}, Nimish P. and {Holwerda}, Benne W. and {Huertas-Company}, Marc and {Iyer}, Kartheik G. and {Jogee}, Shardha and {Jung}, Intae and {Kewley}, Lisa J. and {Kirkpatrick}, Allison and {Koekemoer}, Anton M. and {Liu}, James and {Lotz}, Jennifer M. and {Lucas}, Ray A. and {Newman}, Jeffrey A. and {Pacifici}, Camilla and {Pandya}, Viraj and {Papovich}, Casey and {Pentericci}, Laura and {P{\'e}rez-Gonz{\'a}lez}, Pablo G. and {Petersen}, Jayse and {Pirzkal}, Nor and {Rafelski}, Marc and {Ravindranath}, Swara and {Simons}, Raymond C. and {Snyder}, Gregory F. and {Somerville}, Rachel S. and {Stanway}, Elizabeth R. and {Straughn}, Amber N. and {Tacchella}, Sandro and {Trump}, Jonathan R. and {Vega-Ferrero}, Jes{\'u}s and {Wilkins}, Stephen M. and {Yang}, Guang and {Zavala}, Jorge A.},
        title = "{CEERS Key Paper. III. The Diversity of Galaxy Structure and Morphology at z = 3-9 with JWST}",
      journal = {\apjl},
         year = 2023,
        month = mar,
       volume = {946},
       number = {1},
          eid = {L15},
        pages = {L15},
          doi = {10.3847/2041-8213/acad01},
archivePrefix = {arXiv},
       eprint = {2210.14713},
 primaryClass = {astro-ph.GA},
       adsurl = {https://ui.adsabs.harvard.edu/abs/2023ApJ...946L..15K}
}

@ARTICLE{Robertson2023,
       author = {{Robertson}, Brant E. and {Tacchella}, Sandro and {Johnson}, Benjamin D. and {Hausen}, Ryan and {Alabi}, Adebusola B. and {Boyett}, Kristan and {Bunker}, Andrew J. and {Carniani}, Stefano and {Egami}, Eiichi and {Eisenstein}, Daniel J. and {Hainline}, Kevin N. and {Helton}, Jakob M. and {Ji}, Zhiyuan and {Kumari}, Nimisha and {Lyu}, Jianwei and {Maiolino}, Roberto and {Nelson}, Erica J. and {Rieke}, Marcia J. and {Shivaei}, Irene and {Sun}, Fengwu and {{\"U}bler}, Hannah and {Williams}, Christina C. and {Willmer}, Christopher N.~A. and {Witstok}, Joris},
        title = "{Morpheus Reveals Distant Disk Galaxy Morphologies with JWST: The First AI/ML Analysis of JWST Images}",
      journal = {\apjl},
         year = 2023,
        month = jan,
       volume = {942},
       number = {2},
          eid = {L42},
        pages = {L42},
          doi = {10.3847/2041-8213/aca086},
archivePrefix = {arXiv},
       eprint = {2208.11456},
 primaryClass = {astro-ph.GA},
       adsurl = {https://ui.adsabs.harvard.edu/abs/2023ApJ...942L..42R}
}

@ARTICLE{Norris1985,
       author = {{Norris}, J. and {Bessell}, M.~S. and {Pickles}, A.~J.},
        title = "{Population studies. I. The Bidelman-MacConnell ``weak-metal'' stars.}",
      journal = {\apjs},
         year = 1985,
        month = jul,
       volume = {58},
        pages = {463-492},
          doi = {10.1086/191049},
       adsurl = {https://ui.adsabs.harvard.edu/abs/1985ApJS...58..463N}
}

@ARTICLE{Chiba2000,
       author = {{Chiba}, Masashi and {Beers}, Timothy C.},
        title = "{Kinematics of Metal-poor Stars in the Galaxy. III. Formation of the Stellar Halo and Thick Disk as Revealed from a Large Sample of Nonkinematically Selected Stars}",
      journal = {\aj},
         year = 2000,
        month = jun,
       volume = {119},
       number = {6},
        pages = {2843-2865},
          doi = {10.1086/301409},
archivePrefix = {arXiv},
       eprint = {astro-ph/0003087},
 primaryClass = {astro-ph},
       adsurl = {https://ui.adsabs.harvard.edu/abs/2000AJ....119.2843C}
}

@ARTICLE{Beers2002,
       author = {{Beers}, Timothy C. and {Drilling}, John S. and {Rossi}, Silvia and {Chiba}, Masashi and {Rhee}, Jaehyon and {F{\"u}hrmeister}, Birgit and {Norris}, John E. and {von Hippel}, Ted},
        title = "{Metal Abundances and Kinematics of Bright Metal-poor Giants Selected from the LSE Survey: Implications for the Metal-weak Thick Disk}",
      journal = {\aj},
         year = 2002,
        month = aug,
       volume = {124},
       number = {2},
        pages = {931-948},
          doi = {10.1086/341377},
archivePrefix = {arXiv},
       eprint = {astro-ph/0204339},
 primaryClass = {astro-ph},
       adsurl = {https://ui.adsabs.harvard.edu/abs/2002AJ....124..931B}
}

@ARTICLE{Ruchti2010,
       author = {{Ruchti}, G.~R. and {Fulbright}, J.~P. and {Wyse}, R.~F.~G. and {Gilmore}, G.~F. and {Bienaym{\'e}}, O. and {Binney}, J. and {Bland-Hawthorn}, J. and {Campbell}, R. and {Freeman}, K.~C. and {Gibson}, B.~K. and {Grebel}, E.~K. and {Helmi}, A. and {Munari}, U. and {Navarro}, J.~F. and {Parker}, Q.~A. and {Reid}, W. and {Seabroke}, G.~M. and {Siebert}, A. and {Siviero}, A. and {Steinmetz}, M. and {Watson}, F.~G. and {Williams}, M. and {Zwitter}, T.},
        title = "{Origins of the Thick Disk as Traced by the Alpha Elements of Metal-poor Giant Stars Selected from Rave}",
      journal = {\apjl},
         year = 2010,
        month = oct,
       volume = {721},
       number = {2},
        pages = {L92-L96},
          doi = {10.1088/2041-8205/721/2/L92},
archivePrefix = {arXiv},
       eprint = {1008.3828},
 primaryClass = {astro-ph.GA},
       adsurl = {https://ui.adsabs.harvard.edu/abs/2010ApJ...721L..92R}
}

@ARTICLE{Beers2014,
       author = {{Beers}, Timothy C. and {Norris}, John E. and {Placco}, Vinicius M. and {Lee}, Young Sun and {Rossi}, Silvia and {Carollo}, Daniela and {Masseron}, Thomas},
        title = "{Population Studies. XIII. A New Analysis of the Bidelman-MacConnell ``Weak-metal'' Stars{\textemdash}Confirmation of Metal-poor Stars in the Thick Disk of the Galaxy}",
      journal = {\apj},
         year = 2014,
        month = oct,
       volume = {794},
       number = {1},
          eid = {58},
        pages = {58},
          doi = {10.1088/0004-637X/794/1/58},
archivePrefix = {arXiv},
       eprint = {1408.3165},
 primaryClass = {astro-ph.GA},
       adsurl = {https://ui.adsabs.harvard.edu/abs/2014ApJ...794...58B}
}

@ARTICLE{Santistevan2021,
       author = {{Santistevan}, Isaiah B. and {Wetzel}, Andrew and {Sanderson}, Robyn E. and {El-Badry}, Kareem and {Samuel}, Jenna and {Faucher-Gigu{\`e}re}, Claude-Andr{\'e}},
        title = "{The origin of metal-poor stars on prograde disc orbits in FIRE simulations of Milky Way-mass galaxies}",
      journal = {\mnras},
         year = 2021,
        month = jul,
       volume = {505},
       number = {1},
        pages = {921-938},
          doi = {10.1093/mnras/stab1345},
archivePrefix = {arXiv},
       eprint = {2102.03369},
 primaryClass = {astro-ph.GA},
       adsurl = {https://ui.adsabs.harvard.edu/abs/2021MNRAS.505..921S}
}

@ARTICLE{Sestito2021,
       author = {{Sestito}, Federico and {Buck}, Tobias and {Starkenburg}, Else and {Martin}, Nicolas F. and {Navarro}, Julio F. and {Venn}, Kim A. and {Obreja}, Aura and {Jablonka}, Pascale and {Macci{\`o}}, Andrea V.},
        title = "{Exploring the origin of low-metallicity stars in Milky-Way-like galaxies with the NIHAO-UHD simulations}",
      journal = {\mnras},
         year = 2021,
        month = jan,
       volume = {500},
       number = {3},
        pages = {3750-3762},
          doi = {10.1093/mnras/staa3479},
archivePrefix = {arXiv},
       eprint = {2009.14207},
 primaryClass = {astro-ph.GA},
       adsurl = {https://ui.adsabs.harvard.edu/abs/2021MNRAS.500.3750S}
}

@ARTICLE{SotilloRamos2023,
       author = {{Sotillo-Ramos}, Diego and {Bergemann}, Maria and {Friske}, Jennifer K.~S. and {Pillepich}, Annalisa},
        title = "{On the likelihoods of finding very metal-poor (and old) stars in the Milky Way's disc, bulge, and halo}",
      journal = {\mnras},
         year = 2023,
        month = oct,
       volume = {525},
       number = {1},
        pages = {L105-L111},
          doi = {10.1093/mnrasl/slad103},
archivePrefix = {arXiv},
       eprint = {2307.14421},
 primaryClass = {astro-ph.GA},
       adsurl = {https://ui.adsabs.harvard.edu/abs/2023MNRAS.525L.105S}
}

@ARTICLE{McCluskey2024,
       author = {{McCluskey}, Fiona and {Wetzel}, Andrew and {Loebman}, Sarah R. and {Moreno}, Jorge and {Faucher-Gigu{\`e}re}, Claude-Andr{\'e} and {Hopkins}, Philip F.},
        title = "{Disc settling and dynamical heating: histories of Milky Way-mass stellar discs across cosmic time in the FIRE simulations}",
      journal = {\mnras},
         year = 2024,
        month = jan,
       volume = {527},
       number = {3},
        pages = {6926-6949},
          doi = {10.1093/mnras/stad3547},
archivePrefix = {arXiv},
       eprint = {2303.14210},
 primaryClass = {astro-ph.GA},
       adsurl = {https://ui.adsabs.harvard.edu/abs/2024MNRAS.527.6926M}
}

@ARTICLE{Randich2022,
       author = {{Randich}, S. and {Gilmore}, G. and {Magrini}, L. and {Sacco}, G.~G. and {Jackson}, R.~J. and {Jeffries}, R.~D. and {Worley}, C.~C. and {Hourihane}, A. and {Gonneau}, A. and {Viscasillas Vazquez}, C. and {Franciosini}, E. and {Lewis}, J.~R. and {Alfaro}, E.~J. and {Allende Prieto}, C. and {Bensby}, T. and {Blomme}, R. and {Bragaglia}, A. and {Flaccomio}, E. and {Fran{\c{c}}ois}, P. and {Irwin}, M.~J. and {Koposov}, S.~E. and {Korn}, A.~J. and {Lanzafame}, A.~C. and {Pancino}, E. and {Recio-Blanco}, A. and {Smiljanic}, R. and {Van Eck}, S. and {Zwitter}, T. and {Asplund}, M. and {Bonifacio}, P. and {Feltzing}, S. and {Binney}, J. and {Drew}, J. and {Ferguson}, A.~M.~N. and {Micela}, G. and {Negueruela}, I. and {Prusti}, T. and {Rix}, H.-W. and {Vallenari}, A. and {Bayo}, A. and {Bergemann}, M. and {Biazzo}, K. and {Carraro}, G. and {Casey}, A.~R. and {Damiani}, F. and {Frasca}, A. and {Heiter}, U. and {Hill}, V. and {Jofr{\'e}}, P. and {de Laverny}, P. and {Lind}, K. and {Marconi}, G. and {Martayan}, C. and {Masseron}, T. and {Monaco}, L. and {Morbidelli}, L. and {Prisinzano}, L. and {Sbordone}, L. and {Sousa}, S.~G. and {Zaggia}, S. and {Adibekyan}, V. and {Bonito}, R. and {Caffau}, E. and {Daflon}, S. and {Feuillet}, D.~K. and {Gebran}, M. and {Gonzalez Hernandez}, J.~I. and {Guiglion}, G. and {Herrero}, A. and {Lobel}, A. and {Maiz Apellaniz}, J. and {Merle}, T. and {Mikolaitis}, {\v{S}}. and {Montes}, D. and {Morel}, T. and {Soubiran}, C. and {Spina}, L. and {Tabernero}, H.~M. and {Tautvai{\v{s}}iene}, G. and {Traven}, G. and {Valentini}, M. and {Van der Swaelmen}, M. and {Villanova}, S. and {Wright}, N.~J. and {Abbas}, U. and {Aguirre B{\o}rsen-Koch}, V. and {Alves}, J. and {Balaguer-Nunez}, L. and {Barklem}, P.~S. and {Barrado}, D. and {Berlanas}, S.~R. and {Binks}, A.~S. and {Bressan}, A. and {Capuzzo-Dolcetta}, R. and {Casagrande}, L. and {Casamiquela}, L. and {Collins}, R.~S. and {D'Orazi}, V. and {Dantas}, M.~L.~L. and {Debattista}, V.~P. and {Delgado-Mena}, E. and {Di Marcantonio}, P. and {Drazdauskas}, A. and {Evans}, N.~W. and {Famaey}, B. and {Franchini}, M. and {Fr{\'e}mat}, Y. and {Friel}, E.~D. and {Fu}, X. and {Geisler}, D. and {Gerhard}, O. and {Gonzalez Solares}, E.~A. and {Grebel}, E.~K. and {Gutierrez Albarran}, M.~L. and {Hatzidimitriou}, D. and {Held}, E.~V. and {Jim{\'e}nez-Esteban}, F. and {J{\"o}nsson}, H. and {Jordi}, C. and {Khachaturyants}, T. and {Kordopatis}, G. and {Kos}, J. and {Lagarde}, N. and {Mahy}, L. and {Mapelli}, M. and {Marfil}, E. and {Martell}, S.~L. and {Messina}, S. and {Miglio}, A. and {Minchev}, I. and {Moitinho}, A. and {Montalban}, J. and {Monteiro}, M.~J.~P.~F.~G. and {Morossi}, C. and {Mowlavi}, N. and {Mucciarelli}, A. and {Murphy}, D.~N.~A. and {Nardetto}, N. and {Ortolani}, S. and {Paletou}, F. and {Palou{\v{s}}}, J. and {Paunzen}, E. and {Pickering}, J.~C. and {Quirrenbach}, A. and {Re Fiorentin}, P. and {Read}, J.~I. and {Romano}, D. and {Ryde}, N. and {Sanna}, N. and {Santos}, W. and {Seabroke}, G.~M. and {Spagna}, A. and {Steinmetz}, M. and {Stonkut{\'e}}, E. and {Sutorius}, E. and {Th{\'e}venin}, F. and {Tosi}, M. and {Tsantaki}, M. and {Vink}, J.~S. and {Wright}, N. and {Wyse}, R.~F.~G. and {Zoccali}, M. and {Zorec}, J. and {Zucker}, D.~B. and {Walton}, N.~A.},
        title = "{The Gaia-ESO Public Spectroscopic Survey: Implementation, data products, open cluster survey, science, and legacy}",
      journal = {\aap},
         year = 2022,
        month = oct,
       volume = {666},
          eid = {A121},
        pages = {A121},
          doi = {10.1051/0004-6361/202243141},
archivePrefix = {arXiv},
       eprint = {2206.02901},
 primaryClass = {astro-ph.GA},
       adsurl = {https://ui.adsabs.harvard.edu/abs/2022A&A...666A.121R}
}

@ARTICLE{Gilmore2022,
       author = {{Gilmore}, G. and {Randich}, S. and {Worley}, C.~C. and {Hourihane}, A. and {Gonneau}, A. and {Sacco}, G.~G. and {Lewis}, J.~R. and {Magrini}, L. and {Fran{\c{c}}ois}, P. and {Jeffries}, R.~D. and {Koposov}, S.~E. and {Bragaglia}, A. and {Alfaro}, E.~J. and {Allende Prieto}, C. and {Blomme}, R. and {Korn}, A.~J. and {Lanzafame}, A.~C. and {Pancino}, E. and {Recio-Blanco}, A. and {Smiljanic}, R. and {Van Eck}, S. and {Zwitter}, T. and {Bensby}, T. and {Flaccomio}, E. and {Irwin}, M.~J. and {Franciosini}, E. and {Morbidelli}, L. and {Damiani}, F. and {Bonito}, R. and {Friel}, E.~D. and {Vink}, J.~S. and {Prisinzano}, L. and {Abbas}, U. and {Hatzidimitriou}, D. and {Held}, E.~V. and {Jordi}, C. and {Paunzen}, E. and {Spagna}, A. and {Jackson}, R.~J. and {Ma{\'\i}z Apell{\'a}niz}, J. and {Asplund}, M. and {Bonifacio}, P. and {Feltzing}, S. and {Binney}, J. and {Drew}, J. and {Ferguson}, A.~M.~N. and {Micela}, G. and {Negueruela}, I. and {Prusti}, T. and {Rix}, H.-W. and {Vallenari}, A. and {Bergemann}, M. and {Casey}, A.~R. and {de Laverny}, P. and {Frasca}, A. and {Hill}, V. and {Lind}, K. and {Sbordone}, L. and {Sousa}, S.~G. and {Adibekyan}, V. and {Caffau}, E. and {Daflon}, S. and {Feuillet}, D.~K. and {Gebran}, M. and {Gonzalez Hernandez}, J.~I. and {Guiglion}, G. and {Herrero}, A. and {Lobel}, A. and {Merle}, T. and {Mikolaitis}, {\v{S}}. and {Montes}, D. and {Morel}, T. and {Ruchti}, G. and {Soubiran}, C. and {Tabernero}, H.~M. and {Tautvai{\v{s}}ien{\.{e}}}, G. and {Traven}, G. and {Valentini}, M. and {Van der Swaelmen}, M. and {Villanova}, S. and {Viscasillas V{\'a}zquez}, C. and {Bayo}, A. and {Biazzo}, K. and {Carraro}, G. and {Edvardsson}, B. and {Heiter}, U. and {Jofr{\'e}}, P. and {Marconi}, G. and {Martayan}, C. and {Masseron}, T. and {Monaco}, L. and {Walton}, N.~A. and {Zaggia}, S. and {Aguirre B{\o}rsen-Koch}, V. and {Alves}, J. and {Balaguer-Nunez}, L. and {Barklem}, P.~S. and {Barrado}, D. and {Bellazzini}, M. and {Berlanas}, S.~R. and {Binks}, A.~S. and {Bressan}, A. and {Capuzzo-Dolcetta}, R. and {Casagrande}, L. and {Casamiquela}, L. and {Collins}, R.~S. and {D'Orazi}, V. and {Dantas}, M.~L.~L. and {Debattista}, V.~P. and {Delgado-Mena}, E. and {Di Marcantonio}, P. and {Drazdauskas}, A. and {Evans}, N.~W. and {Famaey}, B. and {Franchini}, M. and {Fr{\'e}mat}, Y. and {Fu}, X. and {Geisler}, D. and {Gerhard}, O. and {Gonz{\'a}lez Solares}, E.~A. and {Grebel}, E.~K. and {Guti{\'e}rrez Albarr{\'a}n}, M.~L. and {Jim{\'e}nez-Esteban}, F. and {J{\"o}nsson}, H. and {Khachaturyants}, T. and {Kordopatis}, G. and {Kos}, J. and {Lagarde}, N. and {Ludwig}, H.-G. and {Mahy}, L. and {Mapelli}, M. and {Marfil}, E. and {Martell}, S.~L. and {Messina}, S. and {Miglio}, A. and {Minchev}, I. and {Moitinho}, A. and {Montalban}, J. and {Monteiro}, M.~J.~P.~F.~G. and {Morossi}, C. and {Mowlavi}, N. and {Mucciarelli}, A. and {Murphy}, D.~N.~A. and {Nardetto}, N. and {Ortolani}, S. and {Paletou}, F. and {Palou{\v{s}}}, J. and {Pickering}, J.~C. and {Quirrenbach}, A. and {Re Fiorentin}, P. and {Read}, J.~I. and {Romano}, D. and {Ryde}, N. and {Sanna}, N. and {Santos}, W. and {Seabroke}, G.~M. and {Spina}, L. and {Steinmetz}, M. and {Stonkut{\'e}}, E. and {Sutorius}, E. and {Th{\'e}venin}, F. and {Tosi}, M. and {Tsantaki}, M. and {Wright}, N. and {Wyse}, R.~F.~G. and {Zoccali}, M. and {Zorec}, J. and {Zucker}, D.~B.},
        title = "{The Gaia-ESO Public Spectroscopic Survey: Motivation, implementation, GIRAFFE data processing, analysis, and final data products}",
      journal = {\aap},
         year = 2022,
        month = oct,
       volume = {666},
          eid = {A120},
        pages = {A120},
          doi = {10.1051/0004-6361/202243134},
archivePrefix = {arXiv},
       eprint = {2208.05432},
 primaryClass = {astro-ph.SR},
       adsurl = {https://ui.adsabs.harvard.edu/abs/2022A&A...666A.120G}
}

@ARTICLE{Howes2019,
       author = {{Howes}, Louise M. and {Lindegren}, Lennart and {Feltzing}, Sofia and {Church}, Ross P. and {Bensby}, Thomas},
        title = "{Estimating stellar ages and metallicities from parallaxes and broadband photometry: successes and shortcomings}",
      journal = {\aap},
         year = 2019,
        month = feb,
       volume = {622},
          eid = {A27},
        pages = {A27},
          doi = {10.1051/0004-6361/201833280},
archivePrefix = {arXiv},
       eprint = {1804.08321},
 primaryClass = {astro-ph.SR},
       adsurl = {https://ui.adsabs.harvard.edu/abs/2019A&A...622A..27H}
}

@ARTICLE{Chen2022,
       author = {{Chen}, Xunzhou and {Ge}, Zhishuai and {Chen}, Yuqin and {Bi}, Shaolan and {Yu}, Jie and {Yang}, Wuming and {Ferguson}, Jason W. and {Wu}, Yaqian and {Li}, Yaguang},
        title = "{Ages of Main-sequence Turnoff Stars from the GALAH Survey}",
      journal = {\apj},
         year = 2022,
        month = apr,
       volume = {929},
       number = {2},
          eid = {124},
        pages = {124},
          doi = {10.3847/1538-4357/ac55a1},
archivePrefix = {arXiv},
       eprint = {2203.06083},
 primaryClass = {astro-ph.SR},
       adsurl = {https://ui.adsabs.harvard.edu/abs/2022ApJ...929..124C}
}

@ARTICLE{Viscasillas2023,
       author = {{Viscasillas V{\'a}zquez}, C. and {Magrini}, L. and {Spina}, L. and {Tautvai{\v{s}}ien{\.{e}}}, G. and {Van der Swaelmen}, M. and {Randich}, S. and {Sacco}, G.~G.},
        title = "{The role of radial migration in open cluster and field star populations with Gaia DR3}",
      journal = {\aap},
         year = 2023,
        month = nov,
       volume = {679},
          eid = {A122},
        pages = {A122},
          doi = {10.1051/0004-6361/202346963},
archivePrefix = {arXiv},
       eprint = {2309.17153},
 primaryClass = {astro-ph.GA},
       adsurl = {https://ui.adsabs.harvard.edu/abs/2023A&A...679A.122V}
}

@ARTICLE{Kordopatis2023,
       author = {{Kordopatis}, G. and {Schultheis}, M. and {McMillan}, P.~J. and {Palicio}, P.~A. and {de Laverny}, P. and {Recio-Blanco}, A. and {Creevey}, O. and {{\'A}lvarez}, M.~A. and {Andrae}, R. and {Poggio}, E. and {Spitoni}, E. and {Contursi}, G. and {Zhao}, H. and {Oreshina-Slezak}, I. and {Ordenovic}, C. and {Bijaoui}, A.},
        title = "{Stellar ages, masses, extinctions, and orbital parameters based on spectroscopic parameters of Gaia DR3}",
      journal = {\aap},
         year = 2023,
        month = jan,
       volume = {669},
          eid = {A104},
        pages = {A104},
          doi = {10.1051/0004-6361/202244283},
archivePrefix = {arXiv},
       eprint = {2206.07937},
 primaryClass = {astro-ph.GA},
       adsurl = {https://ui.adsabs.harvard.edu/abs/2023A&A...669A.104K}
}

@ARTICLE{McMillan2017,
       author = {{McMillan}, Paul J.},
        title = "{The mass distribution and gravitational potential of the Milky Way}",
      journal = {\mnras},
         year = 2017,
        month = feb,
       volume = {465},
       number = {1},
        pages = {76-94},
          doi = {10.1093/mnras/stw2759},
archivePrefix = {arXiv},
       eprint = {1608.00971},
 primaryClass = {astro-ph.GA},
       adsurl = {https://ui.adsabs.harvard.edu/abs/2017MNRAS.465...76M}
}

@ARTICLE{Ramirez2014,
       author = {{Ram{\'\i}rez}, I. and {Mel{\'e}ndez}, J. and {Bean}, J. and {Asplund}, M. and {Bedell}, M. and {Monroe}, T. and {Casagrande}, L. and {Schirbel}, L. and {Dreizler}, S. and {Teske}, J. and {Tucci Maia}, M. and {Alves-Brito}, A. and {Baumann}, P.},
        title = "{The Solar Twin Planet Search. I. Fundamental parameters of the stellar sample}",
      journal = {\aap},
         year = 2014,
        month = dec,
       volume = {572},
          eid = {A48},
        pages = {A48},
          doi = {10.1051/0004-6361/201424244},
archivePrefix = {arXiv},
       eprint = {1408.4130},
 primaryClass = {astro-ph.SR},
       adsurl = {https://ui.adsabs.harvard.edu/abs/2014A&A...572A..48R}
}

@ARTICLE{Nissen2010,
       author = {{Nissen}, P.~E. and {Schuster}, W.~J.},
        title = "{Two distinct halo populations in the solar neighborhood. Evidence from stellar abundance ratios and kinematics}",
      journal = {\aap},
         year = 2010,
        month = feb,
       volume = {511},
          eid = {L10},
        pages = {L10},
          doi = {10.1051/0004-6361/200913877},
archivePrefix = {arXiv},
       eprint = {1002.4514},
 primaryClass = {astro-ph.GA},
       adsurl = {https://ui.adsabs.harvard.edu/abs/2010A&A...511L..10N}
}

@ARTICLE{Haywood2013,
       author = {{Haywood}, Misha and {Di Matteo}, Paola and {Lehnert}, Matthew D. and {Katz}, David and {G{\'o}mez}, Ana},
        title = "{The age structure of stellar populations in the solar vicinity. Clues of a two-phase formation history of the Milky Way disk}",
      journal = {\aap},
         year = 2013,
        month = dec,
       volume = {560},
          eid = {A109},
        pages = {A109},
          doi = {10.1051/0004-6361/201321397},
archivePrefix = {arXiv},
       eprint = {1305.4663},
 primaryClass = {astro-ph.GA},
       adsurl = {https://ui.adsabs.harvard.edu/abs/2013A&A...560A.109H}
}

@ARTICLE{Helmi2018,
       author = {{Helmi}, Amina and {Babusiaux}, Carine and {Koppelman}, Helmer H. and {Massari}, Davide and {Veljanoski}, Jovan and {Brown}, Anthony G.~A.},
        title = "{The merger that led to the formation of the Milky Way's inner stellar halo and thick disk}",
      journal = {\nat},
         year = 2018,
        month = oct,
       volume = {563},
       number = {7729},
        pages = {85-88},
          doi = {10.1038/s41586-018-0625-x},
archivePrefix = {arXiv},
       eprint = {1806.06038},
 primaryClass = {astro-ph.GA},
       adsurl = {https://ui.adsabs.harvard.edu/abs/2018Natur.563...85H}
}

@ARTICLE{Belokurov2018,
       author = {{Belokurov}, V. and {Erkal}, D. and {Evans}, N.~W. and {Koposov}, S.~E. and {Deason}, A.~J.},
        title = "{Co-formation of the disc and the stellar halo}",
      journal = {\mnras},
         year = 2018,
        month = jul,
       volume = {478},
       number = {1},
        pages = {611-619},
          doi = {10.1093/mnras/sty982},
archivePrefix = {arXiv},
       eprint = {1802.03414},
 primaryClass = {astro-ph.GA},
       adsurl = {https://ui.adsabs.harvard.edu/abs/2018MNRAS.478..611B}
}

@ARTICLE{Reddy2006,
       author = {{Reddy}, Bacham E. and {Lambert}, David L. and {Allende Prieto}, Carlos},
        title = "{Elemental abundance survey of the Galactic thick disc}",
      journal = {\mnras},
         year = 2006,
        month = apr,
       volume = {367},
       number = {4},
        pages = {1329-1366},
          doi = {10.1111/j.1365-2966.2006.10148.x},
archivePrefix = {arXiv},
       eprint = {astro-ph/0512505},
 primaryClass = {astro-ph},
       adsurl = {https://ui.adsabs.harvard.edu/abs/2006MNRAS.367.1329R}
}

@ARTICLE{Berni2025,
       author = {{Berni}, L. and {Spina}, L. and {Magrini}, L. and {Massari}, D. and {Schiappacasse-Ulloa}, J. and {Giribaldi}, R.~E.},
        title = "{Exploring substructures in the Milky Way halo: Neural networks applied to Gaia and APOGEE DR17}",
      journal = {\aap},
         year = 2025,
        month = aug,
       volume = {700},
          eid = {A160},
        pages = {A160},
          doi = {10.1051/0004-6361/202555272},
archivePrefix = {arXiv},
       eprint = {2507.08074},
 primaryClass = {astro-ph.GA},
       adsurl = {https://ui.adsabs.harvard.edu/abs/2025A&A...700A.160B}
}

@ARTICLE{Vasiliev2019,
       author = {{Vasiliev}, Eugene},
        title = "{AGAMA: action-based galaxy modelling architecture}",
      journal = {\mnras},
         year = 2019,
        month = jan,
       volume = {482},
       number = {2},
        pages = {1525-1544},
          doi = {10.1093/mnras/sty2672},
archivePrefix = {arXiv},
       eprint = {1802.08239},
 primaryClass = {astro-ph.GA},
       adsurl = {https://ui.adsabs.harvard.edu/abs/2019MNRAS.482.1525V}
}

@ARTICLE{Massari2019,
       author = {{Massari}, D. and {Koppelman}, H.~H. and {Helmi}, A.},
        title = "{Origin of the system of globular clusters in the Milky Way}",
      journal = {\aap},
         year = 2019,
        month = oct,
       volume = {630},
          eid = {L4},
        pages = {L4},
          doi = {10.1051/0004-6361/201936135},
archivePrefix = {arXiv},
       eprint = {1906.08271},
 primaryClass = {astro-ph.GA},
       adsurl = {https://ui.adsabs.harvard.edu/abs/2019A&A...630L...4M}
}

@ARTICLE{Belokurov2022,
       author = {{Belokurov}, Vasily and {Kravtsov}, Andrey},
        title = "{From dawn till disc: Milky Way's turbulent youth revealed by the APOGEE+Gaia data}",
      journal = {\mnras},
         year = 2022,
        month = jul,
       volume = {514},
       number = {1},
        pages = {689-714},
          doi = {10.1093/mnras/stac1267},
archivePrefix = {arXiv},
       eprint = {2203.04980},
 primaryClass = {astro-ph.GA},
       adsurl = {https://ui.adsabs.harvard.edu/abs/2022MNRAS.514..689B}
}

@ARTICLE{Xiang2022,
       author = {{Xiang}, Maosheng and {Rix}, Hans-Walter},
        title = "{A time-resolved picture of our Milky Way's early formation history}",
      journal = {\nat},
         year = 2022,
        month = mar,
       volume = {603},
       number = {7902},
        pages = {599-603},
          doi = {10.1038/s41586-022-04496-5},
archivePrefix = {arXiv},
       eprint = {2203.12110},
 primaryClass = {astro-ph.GA},
       adsurl = {https://ui.adsabs.harvard.edu/abs/2022Natur.603..599X}
}

@ARTICLE{Bensby2014,
       author = {{Bensby}, T. and {Feltzing}, S. and {Oey}, M.~S.},
        title = "{Exploring the Milky Way stellar disk. A detailed elemental abundance study of 714 F and G dwarf stars in the solar neighbourhood}",
      journal = {\aap},
         year = 2014,
        month = feb,
       volume = {562},
          eid = {A71},
        pages = {A71},
          doi = {10.1051/0004-6361/201322631},
archivePrefix = {arXiv},
       eprint = {1309.2631},
 primaryClass = {astro-ph.GA},
       adsurl = {https://ui.adsabs.harvard.edu/abs/2014A&A...562A..71B}
}

@ARTICLE{Bensby2003,
       author = {{Bensby}, T. and {Feltzing}, S. and {Lundstr{\"o}m}, I.},
        title = "{Elemental abundance trends in the Galactic thin and thick disks as traced by nearby F and G dwarf stars}",
      journal = {\aap},
         year = 2003,
        month = nov,
       volume = {410},
        pages = {527-551},
          doi = {10.1051/0004-6361:20031213},
       adsurl = {https://ui.adsabs.harvard.edu/abs/2003A&A...410..527B}
}

@ARTICLE{Magrini2023,
       author = {{Magrini}, L. and {Viscasillas V{\'a}zquez}, C. and {Spina}, L. and {Randich}, S. and {Romano}, D. and {Franciosini}, E. and {Recio-Blanco}, A. and {Nordlander}, T. and {D'Orazi}, V. and {Baratella}, M. and {Smiljanic}, R. and {Dantas}, M.~L.~L. and {Pasquini}, L. and {Spitoni}, E. and {Casali}, G. and {Van der Swaelmen}, M. and {Bensby}, T. and {Stonkute}, E. and {Feltzing}, S. and {Sacco}, G.~G. and {Bragaglia}, A. and {Pancino}, E. and {Heiter}, U. and {Biazzo}, K. and {Gilmore}, G. and {Bergemann}, M. and {Tautvai{\v{s}}ien{\.{e}}}, G. and {Worley}, C. and {Hourihane}, A. and {Gonneau}, A. and {Morbidelli}, L.},
        title = "{The Gaia-ESO survey: Mapping the shape and evolution of the radial abundance gradients with open clusters}",
      journal = {\aap},
         year = 2023,
        month = jan,
       volume = {669},
          eid = {A119},
        pages = {A119},
          doi = {10.1051/0004-6361/202244957},
archivePrefix = {arXiv},
       eprint = {2210.15525},
 primaryClass = {astro-ph.GA},
       adsurl = {https://ui.adsabs.harvard.edu/abs/2023A&A...669A.119M}
}

@ARTICLE{Sestito2026,
       author = {{Sestito}, Federico and {Fern{\'a}ndez-Alvar}, Emma and {Brooks}, Rebecca and {Olson}, Emma and {Carigi}, Leticia and {Jofr{\'e}}, Paula and {Silva}, Danielle de Brito and {Eldridge}, Camilla J.~L. and {Vitali}, Sara and {Venn}, Kim A. and {Hill}, Vanessa and {Ardern-Arentsen}, Anke and {Kordopatis}, Georges and {Martin}, Nicolas F. and {Navarro}, Julio F. and {Starkenburg}, Else and {Tissera}, Patricia B. and {Jablonka}, Pascale and {Lardo}, Carmela and {Lucchesi}, Romain and {Buck}, Tobias and {Amayo}, Alexia},
        title = "{An ancient system hidden in the Galactic plane?}",
      journal = {\mnras},
         year = 2026,
        month = may,
       volume = {548},
       number = {2},
          eid = {stag563},
        pages = {stag563},
          doi = {10.1093/mnras/stag563},
archivePrefix = {arXiv},
       eprint = {2409.13813},
 primaryClass = {astro-ph.GA},
       adsurl = {https://ui.adsabs.harvard.edu/abs/2026MNRAS.548ag563S}
}

@ARTICLE{Zhang2024,
       author = {{Zhang}, Hanyuan and {Ardern-Arentsen}, Anke and {Belokurov}, Vasily},
        title = "{On the existence of a very metal-poor disc in the Milky Way}",
      journal = {\mnras},
         year = 2024,
        month = sep,
       volume = {533},
       number = {1},
        pages = {889-907},
          doi = {10.1093/mnras/stae1887},
archivePrefix = {arXiv},
       eprint = {2311.09294},
 primaryClass = {astro-ph.GA},
       adsurl = {https://ui.adsabs.harvard.edu/abs/2024MNRAS.533..889Z}
}

@ARTICLE{Bovy2015,
       author = {{Bovy}, Jo},
        title = "{galpy: A python Library for Galactic Dynamics}",
      journal = {\apjs},
         year = 2015,
        month = feb,
       volume = {216},
       number = {2},
          eid = {29},
        pages = {29},
          doi = {10.1088/0067-0049/216/2/29},
archivePrefix = {arXiv},
       eprint = {1412.3451},
 primaryClass = {astro-ph.GA},
       adsurl = {https://ui.adsabs.harvard.edu/abs/2015ApJS..216...29B}
}

@ARTICLE{BailerJones2021,
       author = {{Bailer-Jones}, C.~A.~L. and {Rybizki}, J. and {Fouesneau}, M. and {Demleitner}, M. and {Andrae}, R.},
        title = "{Estimating Distances from Parallaxes. V. Geometric and Photogeometric Distances to 1.47 Billion Stars in Gaia Early Data Release 3}",
      journal = {\aj},
         year = 2021,
        month = mar,
       volume = {161},
       number = {3},
          eid = {147},
        pages = {147},
          doi = {10.3847/1538-3881/abd806},
archivePrefix = {arXiv},
       eprint = {2012.05220},
 primaryClass = {astro-ph.SR},
       adsurl = {https://ui.adsabs.harvard.edu/abs/2021AJ....161..147B}
}

@ARTICLE{Travaglio2004,
       author = {{Travaglio}, Claudia and {Gallino}, Roberto and {Arnone}, Enrico and {Cowan}, John and {Jordan}, Faith and {Sneden}, Christopher},
        title = "{Galactic Evolution of Sr, Y, And Zr: A Multiplicity of Nucleosynthetic Processes}",
      journal = {\apj},
         year = 2004,
        month = feb,
       volume = {601},
       number = {2},
        pages = {864-884},
          doi = {10.1086/380507},
archivePrefix = {arXiv},
       eprint = {astro-ph/0310189},
 primaryClass = {astro-ph},
       adsurl = {https://ui.adsabs.harvard.edu/abs/2004ApJ...601..864T}
}

@ARTICLE{Bisterzo2014,
       author = {{Bisterzo}, S. and {Travaglio}, C. and {Gallino}, R. and {Wiescher}, M. and {K{\"a}ppeler}, F.},
        title = "{Galactic Chemical Evolution and Solar s-process Abundances: Dependence on the $^{13}$C-pocket Structure}",
      journal = {\apj},
         year = 2014,
        month = may,
       volume = {787},
       number = {1},
          eid = {10},
        pages = {10},
          doi = {10.1088/0004-637X/787/1/10},
archivePrefix = {arXiv},
       eprint = {1403.1764},
 primaryClass = {astro-ph.SR},
       adsurl = {https://ui.adsabs.harvard.edu/abs/2014ApJ...787...10B}
}

@ARTICLE{Nissen2015,
       author = {{Nissen}, P.~E.},
        title = "{High-precision abundances of elements in solar twin stars. Trends with stellar age and elemental condensation temperature}",
      journal = {\aap},
         year = 2015,
        month = jul,
       volume = {579},
          eid = {A52},
        pages = {A52},
          doi = {10.1051/0004-6361/201526269},
archivePrefix = {arXiv},
       eprint = {1504.07598},
 primaryClass = {astro-ph.SR},
       adsurl = {https://ui.adsabs.harvard.edu/abs/2015A&A...579A..52N}
}

@ARTICLE{Tautvaisiene2021,
       author = {{Tautvai{\v{s}}ien{\.{e}}}, G. and {Viscasillas V{\'a}zquez}, C. and {Mikolaitis}, {\v{S}}. and {Stonkut{\.{e}}}, E. and {Minkevi{\v{c}}i{\={u}}t{\.{e}}}, R. and {Drazdauskas}, A. and {Bagdonas}, V.},
        title = "{Abundances of neutron-capture elements in thin- and thick-disc stars in the solar neighbourhood}",
      journal = {\aap},
         year = 2021,
        month = may,
       volume = {649},
          eid = {A126},
        pages = {A126},
          doi = {10.1051/0004-6361/202039979},
archivePrefix = {arXiv},
       eprint = {2103.09778},
 primaryClass = {astro-ph.SR},
       adsurl = {https://ui.adsabs.harvard.edu/abs/2021A&A...649A.126T}
}

@ARTICLE{Viscasillas2022,
       author = {{Viscasillas V{\'a}zquez}, C. and {Magrini}, L. and {Casali}, G. and {Tautvai{\v{s}}ien{\.{e}}}, G. and {Spina}, L. and {Van der Swaelmen}, M. and {Randich}, S. and {Bensby}, T. and {Bragaglia}, A. and {Friel}, E. and {Feltzing}, S. and {Sacco}, G.~G. and {Turchi}, A. and {Jim{\'e}nez-Esteban}, F. and {D'Orazi}, V. and {Delgado-Mena}, E. and {Mikolaitis}, {\v{S}}. and {Drazdauskas}, A. and {Minkevi{\v{c}}i{\={u}}t{\.{e}}}, R. and {Stonkut{\.{e}}}, E. and {Bagdonas}, V. and {Montes}, D. and {Guiglion}, G. and {Baratella}, M. and {Tabernero}, H.~M. and {Gilmore}, G. and {Alfaro}, E. and {Francois}, P. and {Korn}, A. and {Smiljanic}, R. and {Bergemann}, M. and {Franciosini}, E. and {Gonneau}, A. and {Hourihane}, A. and {Worley}, C.~C. and {Zaggia}, S.},
        title = "{The Gaia-ESO survey: Age-chemical-clock relations spatially resolved in the Galactic disc}",
      journal = {\aap},
         year = 2022,
        month = apr,
       volume = {660},
          eid = {A135},
        pages = {A135},
          doi = {10.1051/0004-6361/202142937},
archivePrefix = {arXiv},
       eprint = {2202.04863},
 primaryClass = {astro-ph.GA},
       adsurl = {https://ui.adsabs.harvard.edu/abs/2022A&A...660A.135V}
}

@ARTICLE{Feltzing2017,
       author = {{Feltzing}, Sofia and {Howes}, Louise M. and {McMillan}, Paul J. and {Stonkut{\.{e}}}, Edita},
        title = "{On the metallicity dependence of the [Y/Mg]-age relation for solar-type stars}",
      journal = {\mnras},
         year = 2017,
        month = feb,
       volume = {465},
       number = {1},
        pages = {L109-L113},
          doi = {10.1093/mnrasl/slw209},
archivePrefix = {arXiv},
       eprint = {1610.03852},
 primaryClass = {astro-ph.GA},
       adsurl = {https://ui.adsabs.harvard.edu/abs/2017MNRAS.465L.109F}
}

@ARTICLE{Pakstiene2026,
       author = {{Pak{\v{s}}tien{\.{e}}}, E. and {Tautvai{\v{s}}ien{\.{e}}}, G. and {Bagdonas}, V. and {Kjeldsen}, H. and {Winther}, M.~L. and {Drazdauskas}, A. and {Viscasillas V{\'a}zquez}, C. and {Chorniy}, Y. and {Mikolaitis}, {\v{S}}. and {Minkevi{\v{c}}i{\={u}}te}, R. and {Stonkut{\.{e}}}, E.},
        title = "{Calibration of the [C/N] and [Y/Mg] chemical clocks with asteroseismic ages from the TESS space mission}",
      journal = {\aap},
         year = 2026,
        month = apr,
       volume = {708},
          eid = {A250},
        pages = {A250},
          doi = {10.1051/0004-6361/202658894},
archivePrefix = {arXiv},
       eprint = {2602.21413},
 primaryClass = {astro-ph.SR},
       adsurl = {https://ui.adsabs.harvard.edu/abs/2026A&A...708A.250P}
}

@ARTICLE{Alinder2026,
       author = {{Alinder}, S. and {Bensby}, T. and {McMillan}, P.~J.},
        title = "{Impact of selection criteria on the structural parameters of the Galactic thin and thick discs}",
      journal = {\aap},
         year = 2026,
        month = jul,
       volume = {710},
          eid = {A403},
        pages = {A403},
          doi = {10.1051/0004-6361/202558090},
archivePrefix = {arXiv},
       eprint = {2511.10092},
 primaryClass = {astro-ph.GA},
       adsurl = {https://ui.adsabs.harvard.edu/abs/2026A&A...710A.403A}
}

@ARTICLE{DelgadoMena2019,
       author = {{Delgado Mena}, E. and {Moya}, A. and {Adibekyan}, V. and {Tsantaki}, M. and {Gonz{\'a}lez Hern{\'a}ndez}, J.~I. and {Israelian}, G. and {Davies}, G.~R. and {Chaplin}, W.~J. and {Sousa}, S.~G. and {Ferreira}, A.~C.~S. and {Santos}, N.~C.},
        title = "{Abundance to age ratios in the HARPS-GTO sample with Gaia DR2. Chemical clocks for a range of [Fe/H]}",
      journal = {\aap},
         year = 2019,
        month = apr,
       volume = {624},
          eid = {A78},
        pages = {A78},
          doi = {10.1051/0004-6361/201834783},
archivePrefix = {arXiv},
       eprint = {1902.02127},
 primaryClass = {astro-ph.SR},
       adsurl = {https://ui.adsabs.harvard.edu/abs/2019A&A...624A..78D}
}

@ARTICLE{Hawkins2015,
       author = {{Hawkins}, K. and {Jofr{\'e}}, P. and {Masseron}, T. and {Gilmore}, G.},
        title = "{Using chemical tagging to redefine the interface of the Galactic disc and halo}",
      journal = {\mnras},
         year = 2015,
        month = oct,
       volume = {453},
       number = {1},
        pages = {758-774},
          doi = {10.1093/mnras/stv1586},
archivePrefix = {arXiv},
       eprint = {1507.03604},
 primaryClass = {astro-ph.GA},
       adsurl = {https://ui.adsabs.harvard.edu/abs/2015MNRAS.453..758H}
}

@ARTICLE{Das2020,
       author = {{Das}, Payel and {Hawkins}, Keith and {Jofr{\'e}}, Paula},
        title = "{Ages and kinematics of chemically selected, accreted Milky Way halo stars}",
      journal = {\mnras},
         year = 2020,
        month = apr,
       volume = {493},
       number = {4},
        pages = {5195-5207},
          doi = {10.1093/mnras/stz3537},
archivePrefix = {arXiv},
       eprint = {1903.09320},
 primaryClass = {astro-ph.GA},
       adsurl = {https://ui.adsabs.harvard.edu/abs/2020MNRAS.493.5195D}
}

@ARTICLE{Horta2021,
       author = {{Horta}, Danny and {Schiavon}, Ricardo P. and {Mackereth}, J. Ted and {Pfeffer}, Joel and {Mason}, Andrew C. and {Kisku}, Shobhit and {Fragkoudi}, Francesca and {Allende Prieto}, Carlos and {Cunha}, Katia and {Hasselquist}, Sten and {Holtzman}, Jon and {Majewski}, Steven R. and {Nataf}, David and {O'Connell}, Robert W. and {Schultheis}, Mathias and {Smith}, Verne V.},
        title = "{Evidence from APOGEE for the presence of a major building block of the halo buried in the inner Galaxy}",
      journal = {\mnras},
         year = 2021,
        month = jan,
       volume = {500},
       number = {1},
        pages = {1385-1403},
          doi = {10.1093/mnras/staa2987},
archivePrefix = {arXiv},
       eprint = {2007.10374},
 primaryClass = {astro-ph.GA},
       adsurl = {https://ui.adsabs.harvard.edu/abs/2021MNRAS.500.1385H}
}

@ARTICLE{Vasini2024,
       author = {{Vasini}, A. and {Spitoni}, E. and {Matteucci}, F.},
        title = "{Galactic archaeology with [Mg/Mn] versus [Al/Fe] abundance ratios. Uncertainties and caveats}",
      journal = {\aap},
         year = 2024,
        month = mar,
       volume = {683},
          eid = {A121},
        pages = {A121},
          doi = {10.1051/0004-6361/202347603},
archivePrefix = {arXiv},
       eprint = {2310.04530},
 primaryClass = {astro-ph.GA},
       adsurl = {https://ui.adsabs.harvard.edu/abs/2024A&A...683A.121V}
}

@ARTICLE{Kobayashi2006,
       author = {{Kobayashi}, Chiaki and {Umeda}, Hideyuki and {Nomoto}, Ken'ichi and {Tominaga}, Nozomu and {Ohkubo}, Takuya},
        title = "{Galactic Chemical Evolution: Carbon through Zinc}",
      journal = {\apj},
         year = 2006,
        month = dec,
       volume = {653},
       number = {2},
        pages = {1145-1171},
          doi = {10.1086/508914},
archivePrefix = {arXiv},
       eprint = {astro-ph/0608688},
 primaryClass = {astro-ph},
       adsurl = {https://ui.adsabs.harvard.edu/abs/2006ApJ...653.1145K}
}

@ARTICLE{Mishenina2015,
       author = {{Mishenina}, T. and {Gorbaneva}, T. and {Pignatari}, M. and {Thielemann}, F.-K. and {Korotin}, S.~A.},
        title = "{Mn abundances in the stars of the Galactic disc with metallicities -1.0 < [Fe/H] < 0.3}",
      journal = {\mnras},
         year = 2015,
        month = dec,
       volume = {454},
       number = {2},
        pages = {1585-1594},
          doi = {10.1093/mnras/stv2038},
archivePrefix = {arXiv},
       eprint = {1509.05341},
 primaryClass = {astro-ph.GA},
       adsurl = {https://ui.adsabs.harvard.edu/abs/2015MNRAS.454.1585M}
}

@ARTICLE{kim2002,
   author = {{Kim}, Y.-C. and {Demarque}, P. and {Yi}, S.~K. and {Alexander}, D.~R.
	},
    title = "{The Y$^{2}$ Isochrones for {$\alpha$}-Element Enhanced Mixtures}",
  journal = {ApJS},
   eprint = {astro-ph/0208175},
     year = 2002,
    month = dec,
   volume = 143,
    pages = {499-511},
      doi = {10.1086/343041},
   adsurl = {http://adsabs.harvard.edu/abs/2002ApJS..143..499K}
}

@ARTICLE{yi2003,
   author = {{Yi}, S.~K. and {Kim}, Y.-C. and {Demarque}, P.},
    title = "{The Y$^{2}$ Stellar Evolutionary Tracks}",
  journal = {ApJS},
   eprint = {astro-ph/0210201},
     year = 2003,
    month = feb,
   volume = 144,
    pages = {259-261},
      doi = {10.1086/345101},
   adsurl = {http://adsabs.harvard.edu/abs/2003ApJS..144..259Y}
}

@ARTICLE{Lindegren2021A&A...649A...4L,
       author = {{Lindegren}, L. and {Bastian}, U. and {Biermann}, M. and {Bombrun}, A. and {de Torres}, A. and {Gerlach}, E. and {Geyer}, R. and {Hern{\'a}ndez}, J. and {Hilger}, T. and {Hobbs}, D. and {Klioner}, S.~A. and {Lammers}, U. and {McMillan}, P.~J. and {Ramos-Lerate}, M. and {Steidelm{\"u}ller}, H. and {Stephenson}, C.~A. and {van Leeuwen}, F.},
        title = "{Gaia Early Data Release 3. Parallax bias versus magnitude, colour, and position}",
      journal = {\aap},
         year = 2021,
        month = may,
       volume = {649},
          eid = {A4},
        pages = {A4},
          doi = {10.1051/0004-6361/202039653},
archivePrefix = {arXiv},
       eprint = {2012.01742},
 primaryClass = {astro-ph.IM},
       adsurl = {https://ui.adsabs.harvard.edu/abs/2021A&A...649A...4L}
}

@ARTICLE{melendez2012A&A...543A..29M,
       author = {{Mel{\'e}ndez}, J. and {Bergemann}, M. and {Cohen}, J.~G. and {Endl}, M. and {Karakas}, A.~I. and {Ram{\'\i}rez}, I. and {Cochran}, W.~D. and {Yong}, D. and {MacQueen}, P.~J. and {Kobayashi}, C. and {Asplund}, M.},
        title = "{The remarkable solar twin HIP 56948: a prime target in the quest for other Earths}",
      journal = {\aap},
         year = 2012,
        month = jul,
       volume = {543},
          eid = {A29},
        pages = {A29},
          doi = {10.1051/0004-6361/201117222},
archivePrefix = {arXiv},
       eprint = {1204.2766},
 primaryClass = {astro-ph.SR},
       adsurl = {https://ui.adsabs.harvard.edu/abs/2012A&A...543A..29M}
}

@ARTICLE{dotter2017ApJ...840...99D,
       author = {{Dotter}, Aaron and {Conroy}, Charlie and {Cargile}, Phillip and {Asplund}, Martin},
        title = "{The Influence of Atomic Diffusion on Stellar Ages and Chemical Tagging}",
      journal = {\apj},
         year = 2017,
        month = may,
       volume = {840},
       number = {2},
          eid = {99},
        pages = {99},
          doi = {10.3847/1538-4357/aa6d10},
archivePrefix = {arXiv},
       eprint = {1704.03465},
 primaryClass = {astro-ph.SR},
       adsurl = {https://ui.adsabs.harvard.edu/abs/2017ApJ...840...99D}
}

@ARTICLE{green2018JOSS....3..695G,
       author = {{Green}, Gregory M.},
        title = "{dustmaps: A Python interface for maps of interstellar dust}",
      journal = {The Journal of Open Source Software},
         year = 2018,
        month = jun,
       volume = {3},
       number = {26},
        pages = {695},
          doi = {10.21105/joss.00695},
       adsurl = {https://ui.adsabs.harvard.edu/abs/2018JOSS....3..695G}
}

@ARTICLE{sfd1998ApJ...500..525S,
       author = {{Schlegel}, David J. and {Finkbeiner}, Douglas P. and {Davis}, Marc},
        title = "{Maps of Dust Infrared Emission for Use in Estimation of Reddening and Cosmic Microwave Background Radiation Foregrounds}",
      journal = {\apj},
         year = 1998,
        month = jun,
       volume = {500},
       number = {2},
        pages = {525-553},
          doi = {10.1086/305772},
archivePrefix = {arXiv},
       eprint = {astro-ph/9710327},
 primaryClass = {astro-ph},
       adsurl = {https://ui.adsabs.harvard.edu/abs/1998ApJ...500..525S}
}

@ARTICLE{yi2001ApJS..136..417Y,
       author = {{Yi}, Sukyoung and {Demarque}, Pierre and {Kim}, Yong-Cheol and {Lee}, Young-Wook and {Ree}, Chang H. and {Lejeune}, Thibault and {Barnes}, Sydney},
        title = "{Toward Better Age Estimates for Stellar Populations: The Y$^{2}$ Isochrones for Solar Mixture}",
      journal = {\apjs},
         year = 2001,
        month = oct,
       volume = {136},
       number = {2},
        pages = {417-437},
          doi = {10.1086/321795},
archivePrefix = {arXiv},
       eprint = {astro-ph/0104292},
 primaryClass = {astro-ph},
       adsurl = {https://ui.adsabs.harvard.edu/abs/2001ApJS..136..417Y}
}

\clearpage

\let\linenumbers\relax
\let\nolinenumbers\relax
\begin{appendix}

\section{Impact of [$\alpha$/Fe] enhancement on the  determination of age}
\label{app:alpha}

The majority of the stars in our sample have near-solar metallicity. However, since our goal is to find old stars, the most likely candidates are naturally the most metal-poor ones. We applied Eq.~\ref{eq:alpha_enh} to account the effect of [$\alpha$/Fe] on the overall metallicity in our isochrone fitting procedure. The performance of the equation has been tested against isochrones with actual $\alpha$ enhancements by \cite{kim2002}. They found that the equation works remarkably well for [Fe/H] $\sim -2$~dex, precisely reproducing all the isochrone path from the MSTO to the tip of the red giant branch.
On the other hand, they show that for [Fe/H] $\gtrsim -1.5$~dex, only the main sequence is reproducible.
The authors do not recommend the use of the equation for near-solar metallicity because the consistency along the red giant branch becomes highly irregular. 
However, since our sample only consists on MSTO, Eq.~\ref{eq:alpha_enh} can be applied safely. 

\begin{figure}[!htbp]
    \centering
    \includegraphics[width=\columnwidth]{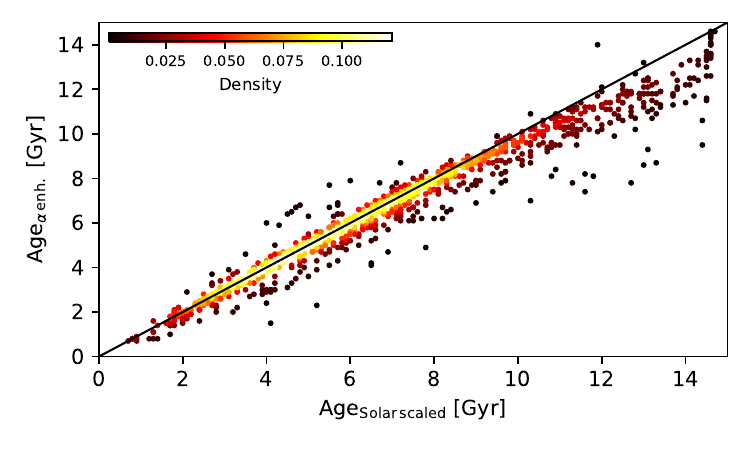}\\
    \includegraphics[width=\columnwidth]{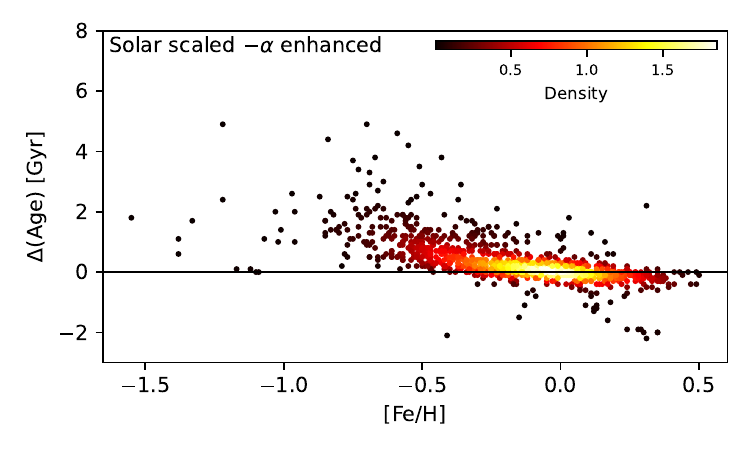}
    \caption{Comparison of age outcomes from isochrones with and without alpha enhancement.
    {\it Top panel:} shows $\alpha$ enhanced determinations via Eq.~\ref{eq:alpha_enh} versus solar scaled determinations. The line of perfect agreement is represented by the diagonal line.
    {\it Bottom panel: } Difference between age determinations as function of metallicity. 
    }
    \label{fig:ages_comp}
\end{figure}

\cite{spina2018MNRAS.474.2580S} have shown experimentally that the $\alpha$ enhancement via Eq.~\ref{eq:alpha_enh} mostly affects the stellar ages grater than $\sim$8~Gyr. However, that  sample consists only on stars of solar-metallicity.
Figure~\ref{fig:ages_comp} shows a comparison using our star sample. Similar to the results of \cite{spina2018MNRAS.474.2580S}, the top panel in the figure displays overestimations from $\alpha$ solar-scaled isochrones for age $\gtrsim 8$~Gyr.
For stars older than $\sim$10~Gyr, ignoring $\alpha$ enhancement frequently yields ages exceeding the age of the universe 13.7~Gyr \citep{plank2020A&A...641A...6P}.
Most of the old stars in the present sample have [Fe/H] $\lesssim -0.4$~dex, as the bottom panel in the figure shows.
For these stars, disregarding $\alpha$ enhancement yields  age overestimations of up to 2~Gyr.

\section{Comparison of reddening estimates}
\label{app:reddening}
We explored the reddening maps available through the \texttt{dustmaps} package \citep{green2018JOSS....3..695G} to identify the most suitable extinction estimates for our stellar sample. Among the available products, {\it Bayestar}  \citep{green2018MNRAS.478..651G} and DECaPS \citep{zucker2025ApJ...992...39Z} were selected as the only large-scale maps providing distance-dependent (3D) reddening estimates, while the map of \cite{sfd1998ApJ...500..525S} (SFD hereafter) was included as the standard all-sky 2D reference. 
SFD estimates were scaled by a factor of 0.86 to account for the systematic overestimation reported by \citet{Schlafly2011ApJ...737..103S}. In addition, to account for the finite distances of our targets, the reddening values were further corrected following the prescription of \citet{Beers2002}; we refer to these values as S\&F+B hereafter.
Other maps were not considered because their spatial coverage or applicability is limited to specific Galactic regions or the local solar neighbourhood, making them unsuitable for our sample.

Only SFD provides reddening estimates for all stars, independently of their sky position. However, since these estimates correspond to the total dust column along the line of sight, applying the correction for the finite distance of the star following \citet{Beers2002} is expected to be appropriate at high Galactic latitudes, in the halo, where the dust distribution is smoother and less affected by the clumpy structure observed near the Galactic plane. The accuracy of this correction may decrease for stars close to the Galaxy plane, where the inhomogeneous distribution of dust becomes standard. 

\begin{figure}[!htbp]
    \centering
    \includegraphics[width=\columnwidth]{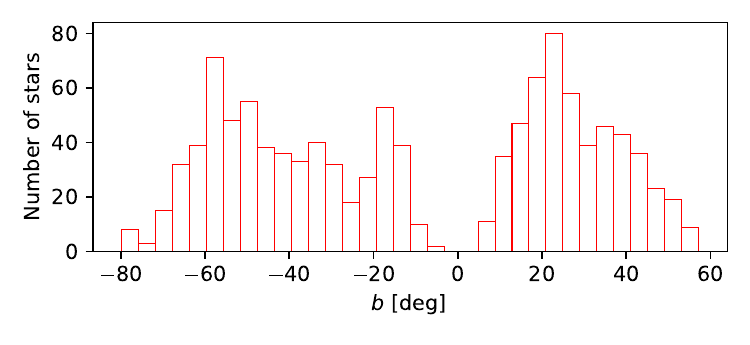}
    \caption{\tiny Histogram of $b$ with bins of $4^\circ$.
    }
    \label{fig:histos}
\end{figure}

Approximately 23\% of our sample is located at low Galactic latitudes ($|b|<20^\circ$, see Fig.~\ref{fig:histos}), where Bayestar should be more suited.
However, bayestar is primarily based on photometry from the Pan-STARRS1 survey \citep{chambers2019panstarrs1surveys}, whose footprint is limited to declinations $\delta \gtrsim -30^\circ$. Consequently, a significant fraction of our southern-sky targets falls outside its coverage. We retrieved Bayestar reddening for $\sim$40\% of our sample (444 stars); these stars will be used below for assessing the relative precision of the S\&F+B values.
The DECaPS footprint, on the other hand, is concentrated towards the southern Galactic plane, covering approximately $240^\circ \lesssim l \lesssim 360^\circ$ and $|b| \lesssim 10^\circ$. Our stellar sample largely avoids this plane region, thus only 16 stars have DECaPS extinction estimates. These stars are used below solely as an independent check of the S\&F+B estimates.

\begin{figure}[!htbp]
    \centering
    \includegraphics[width=\columnwidth]{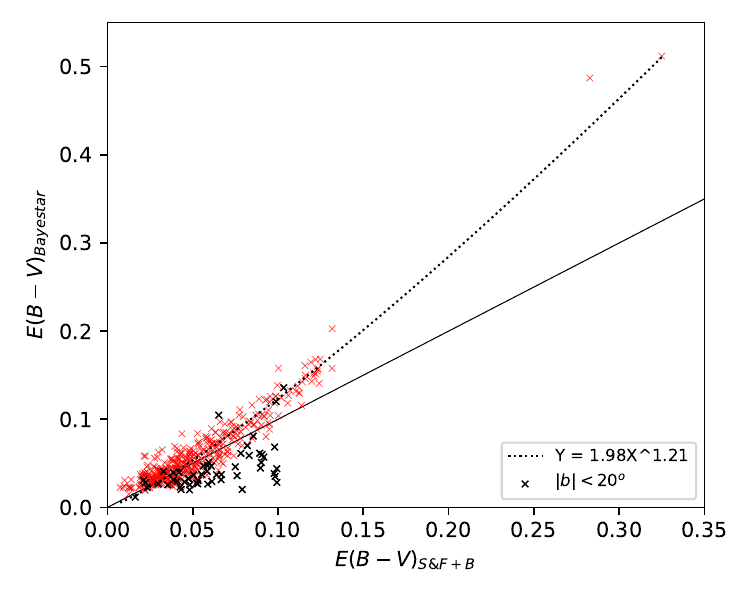}\\
    \includegraphics[width=\columnwidth]{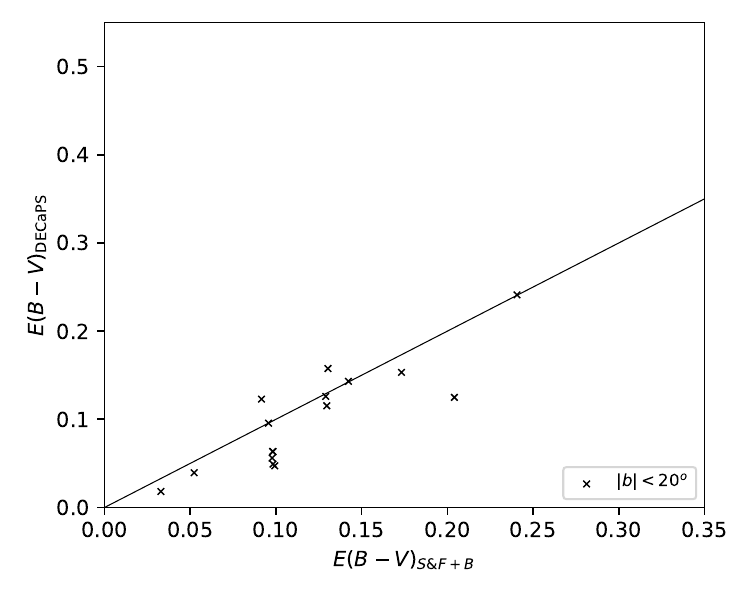}
    \caption{\tiny Comparison between  reddening estimates.
    {\it Top panel:} S\&F+B versus Bayestar. Red and black symbols represent stars with galactic latitude higher and lower than $b = 20$, respectively. The black solid line indicates the prefect agreement. The dotted line represents a power law fitted to the red points, the parameters of which are noted in the legends. {\it Bottom panel: }  S\&F+B versus DECaPS. Symbols are the same as in the top panel.
    }
    \label{fig:ebv}
\end{figure}

Figure~\ref{fig:ebv} shows good agreement between S\&F+B and Bayestar  for $E(B-V)\lesssim0.07$~mag. At higher reddening values, the differences increase systematically, with Bayestar yielding progressively higher values. 
The observed trend is empirically described by a power-law relation, whose parameters are indicated in the plot.
On the other hand, the comparison between S\&F+B and DECaPS seem to show reasonable agreement along $E(B-V)$.
These outcomes may reflect differences in the calibration and methodology of the reddening maps. 
Since the agreement between S\&F+B and DECaPS for $E(B-V) \gtrsim 0.7$ is supported by only a few stars, we adopt the scale of Bayestar. This is, 
we adopt Bayestar estimates when available, and when not, we apply the power law to transform S\&F+B for consistency. 
The vast majority of our sample has S\&F+B $< 0.14$~mag (all but five stars), implying corrections smaller than $+0.03$~mag. Such differences have relatively small impact on the age (i.e. $\lesssim 1.5$~Gyr).
After applying the correction, the standard deviation of the difference between the two sets of values is of $0.013$~mag, which we adopt as the reddening uncertainty.

\begin{figure}[!htbp]
    \centering
    \includegraphics[width=\columnwidth]{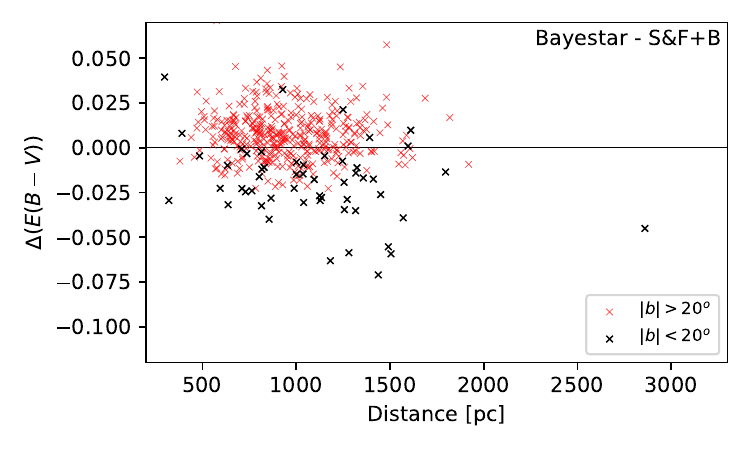}\\
    \includegraphics[width=\columnwidth]{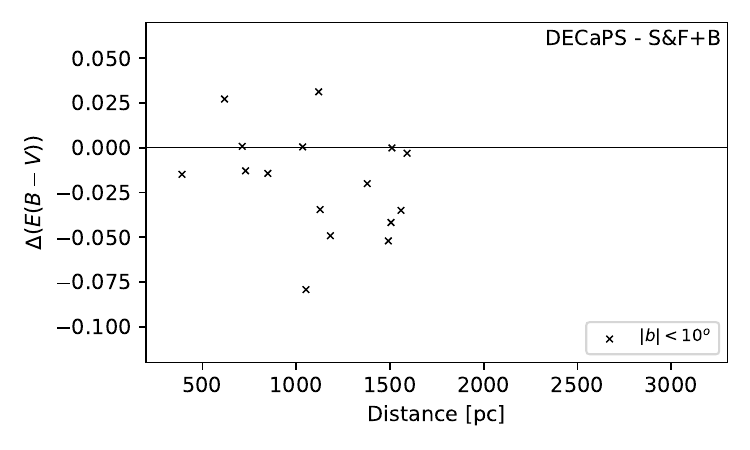}
    \caption{\tiny Comparison between reddening estimates as function of the distance. Difference of survey estimates are given as noted in the plots. The colours of the symbols indicate diverse Galactic latitude ranges as noted in the legends.
    }
    \label{fig:ebv3}
\end{figure}

\begin{figure}[!htbp]
    \centering
    \includegraphics[width=\columnwidth]{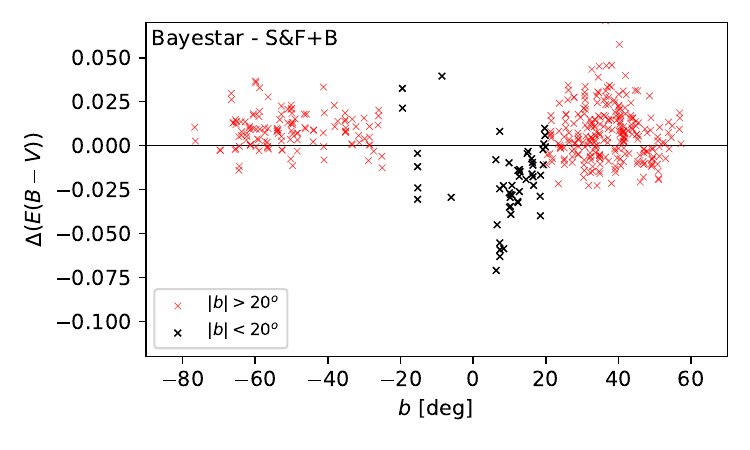}
    \includegraphics[width=\columnwidth]{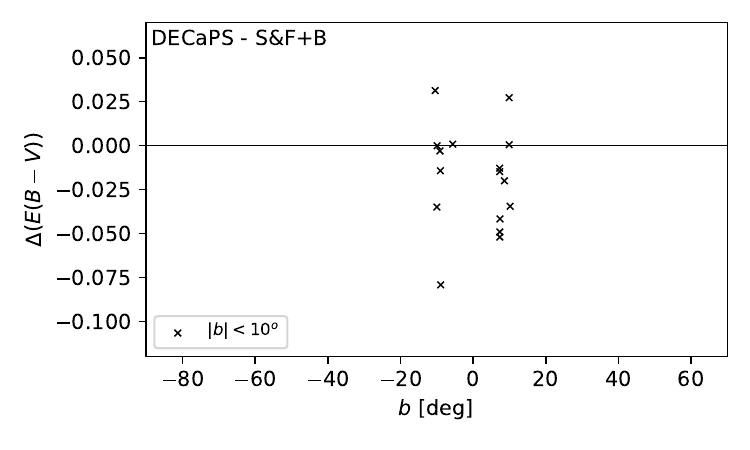}
    \caption{\tiny Comparison between S\&F+B and Bayestar reddening estimates.
    {\it Bottom panel:} Difference between Bayestar reddening and S\&F+B values converted via the equation on the top panel.
    }
    \label{fig:ebv2}
\end{figure}

Figure~\ref{fig:ebv} shows a few outliers in both S\&F+B versus Bayestar and S\&F+B versus DECaPS comparisons, indicating S\&F+B overestimations. These correspond to stars at low Galactic latitudes, and  some  are among the most distant. Figures~\ref{fig:ebv3} and \ref{fig:ebv2} indicate that the stars affected have $|b| < 20^\circ$ and distance $> 1000$~parsec. Our sample has 137 stars (12\%) at that space, among which 105 do not have Bayestar estimates. For them, we simply keep S\&F+B values. These stars are flagged as having likely biased younger ages, which is the effect of overestimating $E(B-V)$.

\FloatBarrier

\section{Age distributions of the disc comparison samples and candidate posterior}
\label{appendix}
The thin-disc comparison sample was selected using both kinematic and chemical information: stars were required to have $V_{\rm tot} < 70~{\rm km~s^{-1}}$ and $Z_{\rm max} < 0.5~{\rm kpc}$, and to satisfy the thin-disc chemical criterion in the $[\alpha/{\rm Fe}]$--$[{\rm Fe/H}]$ plane following \citet{Viscasillas2023}. The thick-disc sample was selected by requiring 
$70 \leq V_{\rm tot} < 180~{\rm km~s^{-1}}$ and 
$Z_{\rm max} > 1~{\rm kpc}$, 
together with the corresponding thick-disc chemical criterion in the 
$[\alpha/{\rm Fe}]$--$[{\rm Fe/H}]$ plane. 
The age distributions of both comparison samples were fitted with Gaussian functions. 
The vertical dashed lines indicate the mean ages of the thin- and thick-disc samples, while the magenta vertical line marks the age of the candidate star.

\begin{figure}[!htbp]
    \centering
    \includegraphics[width=\columnwidth]{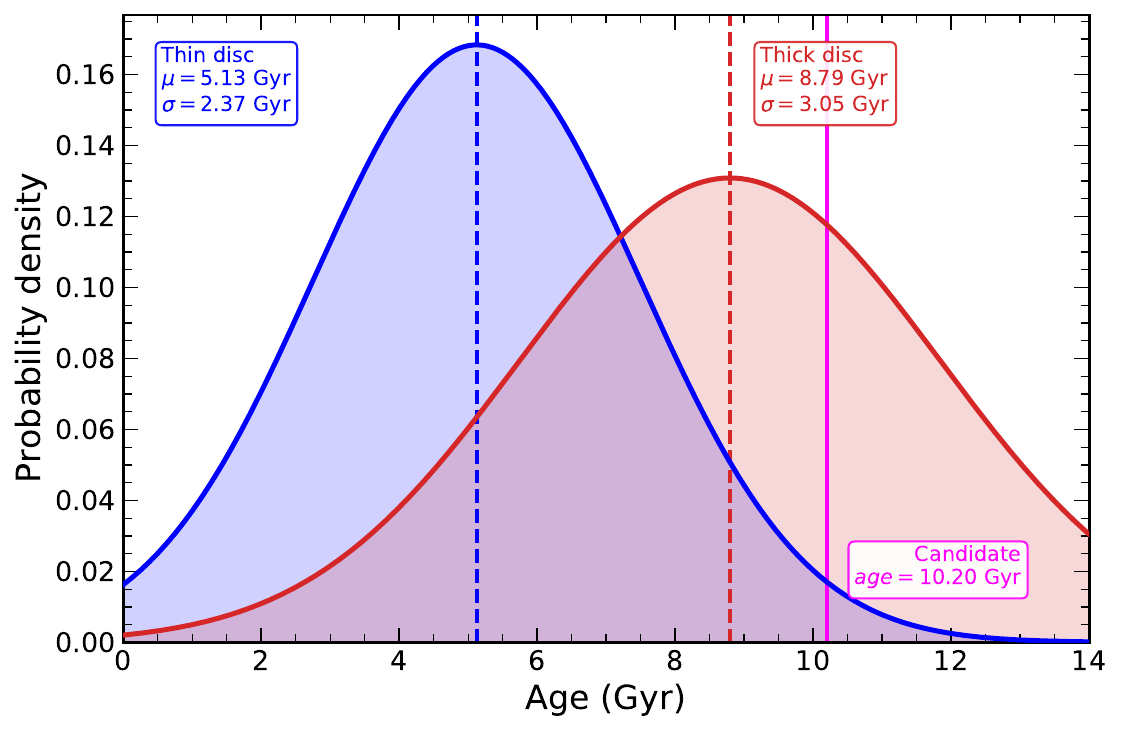}
    \caption{Age distribution comparison between the candidate star and the thin- and thick-disc MSTO samples selected on the basis of their kinematic and chemical properties. 
    The blue and red shaded curves show Gaussian fits to the most probable age distributions of the thin- and thick-disc samples, respectively. 
    The dashed vertical lines indicate the corresponding mean ages, while the magenta vertical line marks the most probable age of the candidate discussed in this work.}
    \label{fig:thin_disc_age_distribution}
\end{figure}

\begin{figure}[!htbp]
    \centering
    \includegraphics[width=\columnwidth]{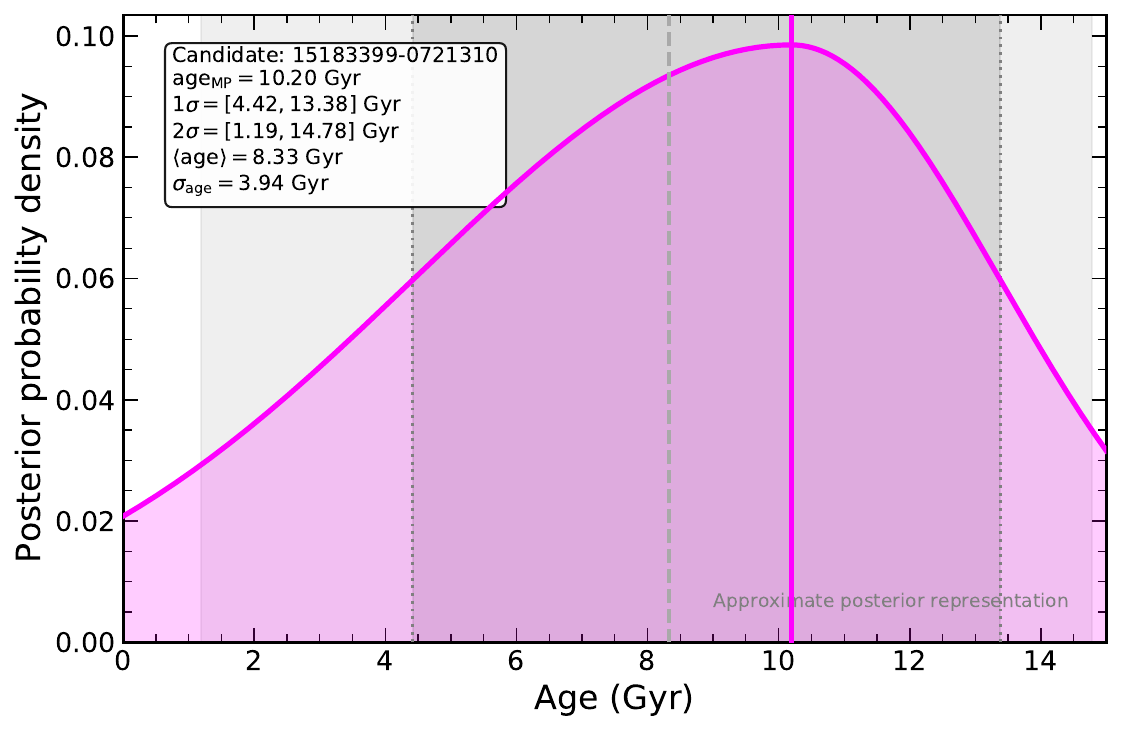}
    \caption{Summary of the age posterior of the candidate star 15183399$-$0721310. The magenta curve shows an approximate representation of the posterior based on the reported most probable age and credible intervals. The solid magenta vertical line marks the most probable age, $\mathrm{age}_{\rm mp}=10.2$ Gyr, while the dashed black line indicates the mean age, $\langle \mathrm{age} \rangle = 8.33$ Gyr. The shaded grey regions show the $1\sigma$-like interval, $\sim 4.42$--$13.38$ Gyr, and the $2\sigma$-like interval, $\sim 1.19$--$14.78$ Gyr.
}
    \label{fig:candidate_age_posterior_summary}
\end{figure}

To further illustrate the age constraint of the candidate star, Fig.~\ref{fig:candidate_age_posterior_summary} shows a summary of its age posterior. The posterior peaks at $\mathrm{age}_{\rm mp}=10.2$ Gyr, but it is broad, with a 1$\sigma$-like interval extending from $\sim 4.42$ to $\sim 13.38$ Gyr and a 2$\sigma$-like interval extending from $\sim 1.19$ to $\sim 14.78$ Gyr. The mean age is $\langle \mathrm{age} \rangle = 8.33$ Gyr, with a standard deviation of $\sigma_{\rm age}=3.94$ Gyr. Thus, the age estimate should be interpreted as indicating that the posterior favours an old solution, rather than as a precise age determination.

\FloatBarrier

\begin{figure}[!htbp]
    \centering
    \includegraphics[width=\columnwidth]{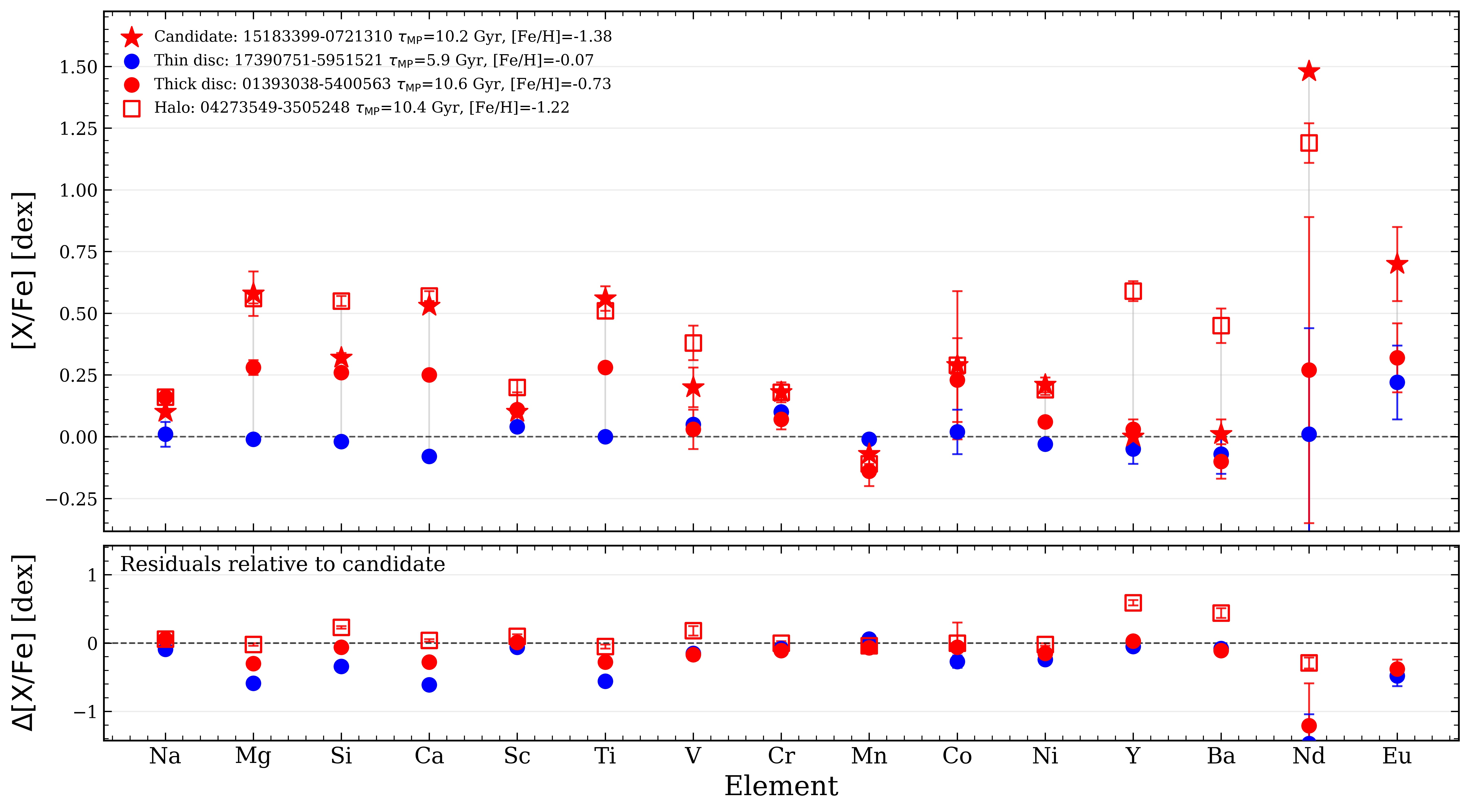}
    \caption{Chemical abundance pattern of the candidate compared with four reference stars. The thin-disc comparison star was selected from the thin-disc subsample defined by combined kinematic and chemical criteria, with age and metallicity close to the typical values of that subsample. The thick-disc star was selected from the corresponding Toomre-defined population, while the halo star was selected from the halo region of the Toomre diagram; in both cases we required ages and metallicities broadly similar to those of the candidate and excluded chemically peculiar abundance patterns. The upper panel shows the abundance ratios $[\mathrm{El/Fe}]$ as a function of element, while the lower panel displays the residuals relative to the candidate.}
    \label{fig:xfe_pattern}
\end{figure}

\FloatBarrier

\section{Kinematics of the chemically selected Mg-enhanced sample}
\label{app:mg_enhanced_kinematics}

\begin{figure}[!htbp]
\centering
\includegraphics[width=\columnwidth]{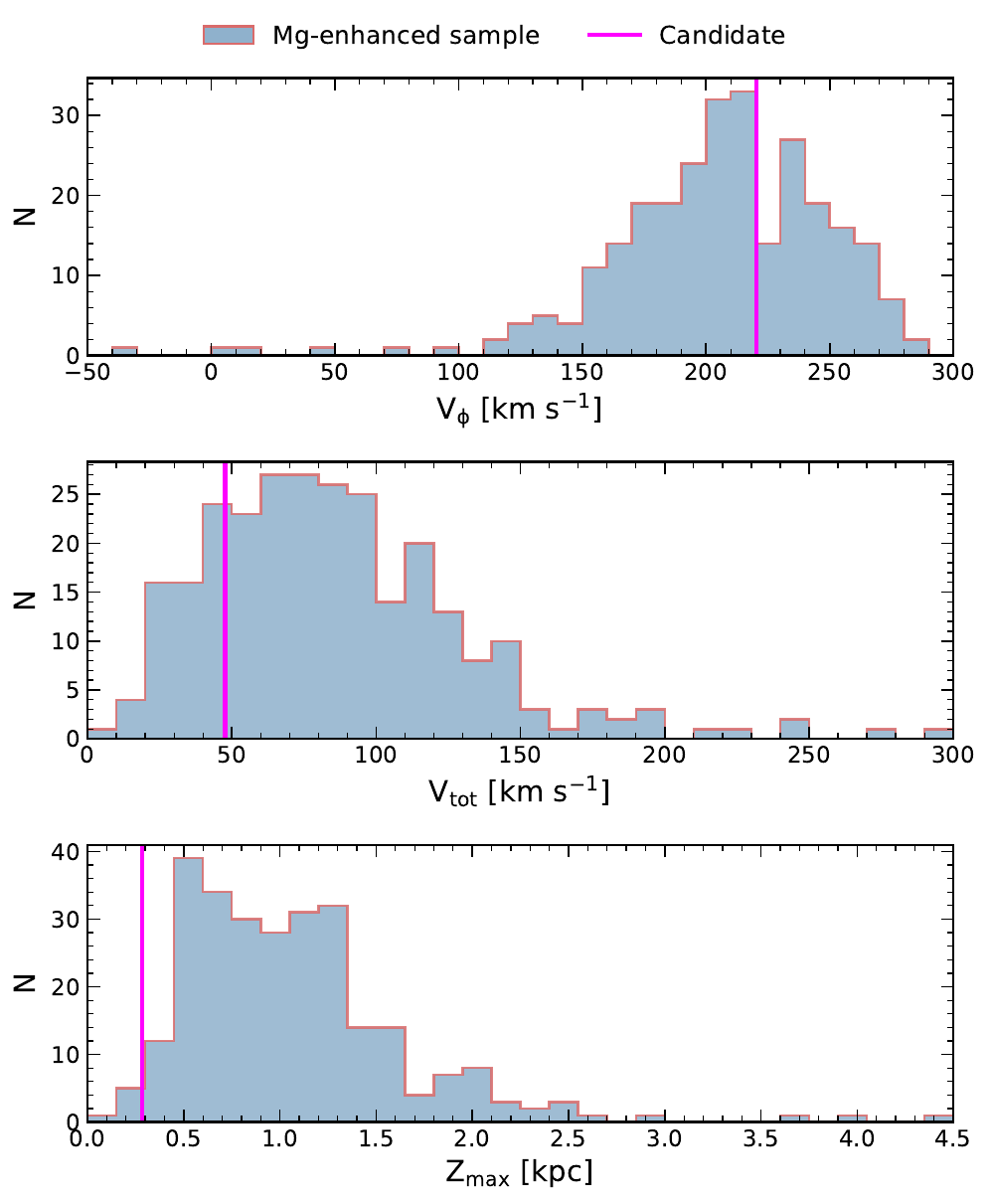}
\caption{Kinematic properties of the chemically selected Mg-enhanced sample. The sample was selected above the chemical boundary in the [Mg/Fe]--[Fe/H] plane, after removing young $\alpha$-rich stars with $\mathrm{age}_{\rm mp}<6$ Gyr; no kinematic cuts were applied. The panels show the distributions of $V_{\phi}$, $V_{\rm tot}$, and $Z_{\max}$. The vertical magenta line marks the position of the candidate.}
\label{fig:mg_enhanced_kinematics}
\end{figure}

To assess whether the candidate could be associated with the MWTD, we examined the kinematic properties of the chemically selected Mg-enhanced population in our sample. The comparison sample was selected above the chemical boundary in the [Mg/Fe]--[Fe/H] plane as defined in section \ref{sec:YMG}, after removing young Mg-rich stars with $\mathrm{age}_{\rm mp}<6$ Gyr, and without applying any kinematic cuts. As shown in Fig. \ref{fig:mg_enhanced_kinematics}, the Mg-enhanced population includes stars with $V_{\phi}$ comparable to the candidate, while the candidate lies at the dynamically colder end of the population in terms of $V_{\rm tot}$ and $Z_{\max}$.

\end{appendix}

\end{document}